\documentclass[conference]{IEEEtran} 
\usepackage{amsmath}
\usepackage{amsmath,amssymb,amsfonts}
\usepackage{cite}
\usepackage{mathtools}
\usepackage[binary-units=true]{siunitx}
\DeclareSIUnit{\belmilliwatt}{Bm}
\usepackage{amsfonts}
\usepackage{amssymb}
\usepackage{graphicx}
\usepackage{multirow}
\usepackage{dsfont}
\usepackage{subfigure}
\usepackage{algorithm}
\usepackage{algpseudocode}
\usepackage{mathtools}

\usepackage{textcomp}
\usepackage[export]{adjustbox}
\usepackage{float}
\usepackage{booktabs}
\usepackage{multicol}
\usepackage{xparse}
\usepackage{color,soul}
\usepackage[bottom]{footmisc}
\usepackage[binary-units=true]{siunitx}
\DeclareSIUnit{\belmilliwatt}{Bm}
\DeclareSIUnit{\dBm}{\deci\belmilliwatt}
\DeclareSIUnit[per-mode=symbol,per-symbol=p]{\Bps}{\byte\per\second}

\sethlcolor{yellow}
\usepackage{algpseudocode}
\DeclareMathOperator{\E}{\mathbb{E}}

\usepackage{hyperref}

\makeatletter
\def\BState{\State\hskip-\ALG@thistlm}
\makeatother

\begin{document}


\title{Distributed Backlog-Aware D2D Communication for Heterogeneous IIoT Applications}

\author{
\IEEEauthorblockN{Hossam Farag\IEEEauthorrefmark{1}, \v{C}edomir Stefanovi\'{c}\IEEEauthorrefmark{1} and Mikael Gidlund\IEEEauthorrefmark{2}}
\IEEEauthorblockA{
\IEEEauthorrefmark{1}Department of Electronic Systems, Aalborg University, Denmark\\
\IEEEauthorrefmark{2}Department of Information Systems and Technology, Mid Sweden University, Sweden \\
Email: hmf@es.aau.dk,cs@es.aau.dk, mikael.gidlund@miun.se}}

	\maketitle
	\begin{abstract}
Delay and Age-of-Information (AoI) are two crucial performance metrics for emerging time-sensitive applications in Industrial Internet of Things (IIoT). In order to achieve optimal performance, studying the inherent interplay between these two parameters in non-trivial task. In this work, we consider a Device-to-Device (D2D)-based heterogeneous IIoT network that supports two types of traffic flows, namely AoI-orientated. First, we introduce a distributed backlog-aware random access protocol that allows the AoI-orientated nodes to opportunistically access the channel based on the queue occupancy of the delay-oriented node. Then, we develop an analytical framework to evaluate the average delay and the average AoI, and formulate an optimization problem to minimize the AoI under a given delay constraint. Finally, we provide numerical results to demonstrate the impact of different network parameters on the performance in terms of the average delay and the average AoI. We also give the numerical solutions of the optimal parameters that minimize the AoI subject to a delay constraint.
	\end{abstract}
\section{Introduction} \label{intro}
The emerging technology of Internet of Things (IoT) enables ubiquitous connectivity and innovative services spanning a wide range of applications, such as industrial automation, healthcare and intelligent transportation~\cite{iot}. The realization of such applications implies the interconnection of a large number of heterogeneous and pervasive IoT devices continuously generating and transmitting sensor data to their respective destinations. According to the Global System for Mobile Communications Association (GSMA), it is expected that the number of connected IoT devices will grow up to 75 billion by 2025~\cite{iot_number}. As a subset of the IoT, Industrial IoT (IIoT) has gained significant attention as a key technology supporting machine type communication for time-sensitive industrial applications~\cite{IIoT}. For a vast number of IIoT applications, the field network comprises a high number of sensors and actuators that sense the physical environment and perform appropriate actions, respectively~\cite{dense_IIoT}. Different from the conventional star-network configuration with a static sink, Device-to-device (D2D) communications provide a promising transmission paradigm for sensor-actuator pairs which are in propinquity~\cite{D2D_IIoT}. In such setups, sensors directly transmit data to actuators without involving a central node, resulting in improved throughput and energy efficiency. 

Typically, the field network in IIoT applications supports transmissions of heterogeneous traffic flows which are characterized by different performance objectives~\cite{Hossam}. In process monitoring and control scenarios, the distributed sensor nodes either report special events to their corresponding actuators that perform actions in response, or transmit either status updates about a certain proces. 
 In the former case, it is essential to maintain low latency communication of event reporting from sensors to actuators in order to avoid production inefficiency or safety-critical situations. An event could be any incident happening, e.g., fire or leakage of gas, that should be delivered to the corresponding actuator within a predefined deadline. 
 In the latter case, the freshness of the status updates is crucial for proper functioning of the system as the actuation decisions are mainly influenced by the freshest information available~\cite{AoI_D2D_IIoT}. For instance, in oil refineries, valve actuators should acquire timely monitoring of oil level to avoid spilling of oil tanks~\cite{oil_refinery}. Classical time-related performance metrics, such as throughput and delay, have a transmitter-side view of the channel and are insufficient to reflect information freshness at the receiver side. Age of Information (AoI) has been proposed to characterize information freshness which measures the time elapsed since the generation of the most recently received packet~\cite{AoI_def}. Although D2D communication can improve throughput and energy efficiency~\cite{D2D_Energy}, the high number of deployed sensor-actuator pairs and the uncoordinated channel access pose a challenge on guaranteeing the performance objectives (low latency and low AoI) of heterogeneous traffic streams.
 
In this work, we consider a D2D-based heterogeneous IIoT network that supports two traffic flows, delay-oriented traffic and AoI-oriented traffic. The delay-oriented traffic is transmitted by a single node and is subject to a predefined delay constraint, while the AoI-oriented traffic is transmitted by a set of nodes distributed according to a stochastic point process. Our contributions in this work can be summarized as follows: 
  \begin{itemize}
  \item  We introduce a distributed backlog-aware random access protocol, in which the AoI-oriented nodes access the shared channel based on the backlog status of the delay-oriented node. 
  
\item  Using stochastic geometry, Discrete-Time Markov Chain (DTMC) and queueing theory, we derive closed-form expressions for the average AoI, average delay, and the average queue size of the delay-oriented node, and formulate an optimization problem to minimize the average AoI under a given delay constraint.

\item We evaluate the network performance using numerical results in order to investigate the impact of different network parameters on the average AoI and the average delay.

\item We solve the formulated optimization via a 2-D exhaustive search under different network parameters.
 \end{itemize}

\section{Related Work}

AoI has been receiving a growing attention as a promising metric for time-sensitive applications. Several works were conducted to study and analyze AoI utilizing queuing theory~\cite{Q2, Q3}. The AoI minimization problem has been addressed through different approaches including packet management schemes~\cite{q_management, management}, adjusting sampling rate~\cite{AoI_def} and introducing packet deadlines~\cite{deadlines}.  The queuing analysis in these works considers only the temporal traffic dynamics associated with the transmitters and receivers based on the assumption of an error-free channel, i.e., without considering transmission failures due to interference and collisions that are inevitable in the industrial environment.
In~\cite{schedule_1, schedule_2, schedule_4, schedule_5}, the authors introduced a number of scheduling protocols with the goal to minimize the average AoI considering wireless networks with unreliable channels.
All these works are based on a centralized scheduling policy which is not applicable to the D2D communication scenario in IIoT, as it would incur high communication overhead and cannot scale with the network size.

The main research on scheduling in wireless D2D networks focuses on the analysis and optimization of resource allocation~\cite{D2D_resource}, traffic density~\cite{D2D_traffic}, or user fairness~\cite{D2D_fairness}, however, there are few works that focus on the study and minimization of AoI 
The authors in in~\cite{D2D_AoI_0} statistically analyzed the peak AoI and the conditional success probability in a D2D communication scenario under preemptive and non-preemptive queues.
Based on a stochastic geometry approach, the authors in~\cite{D2D_AoI_1} proposed a distributed scheduling policy that utilizes local observations, encapsulated in the concept of stopping sets to make scheduling decisions that minimize the network-wide average AoI. The work in~\cite{D2D_AoI_2} introduced a joint optimization problem of the transmit probability and transmit power to maximize the energy efficiency of a covert D2D communication under AoI constraint. The authors in~\cite{D2D_AoI_3} proposed an age-aware scheduling algorithm where edge controllers assist each D2D node to select a proper resource block and  transmit power level to satisfy the AoI constraint. All these works consider that the network supports only a single traffic flow, which is age-oriented traffic and aim to minimize the average AoI without considering the coexistence of other traffic flows. The works in~\cite{schedule_3, D2D_AoI_5} proposed scheduling policies to optimize the AoI while satisfying throughput constraints. The authors in~\cite{D2D_AoI_4} proposed a link scheduling policy for a joint optimization of AoI and throughput based on a deep learning approach. These works assume that, at each timeslot, the same user transmit either AoI-oriented packet or a throughput-oriented packet. The works in~\cite{AoI_Delay_multiserver1, AoI_Delay_multiserver2} study the AoI-delay tradeoff for multi-server queuing systems considering error-free transmissions which is not applicable to real IIoT systems. In~\cite{AoI_Delay_1}, the authors studied the delay and AoI violation probabilities considering finite-block length packets under noisy channels, however, the study was conducted separately for the two parameters.

 \section{System Model}

    \begin{figure}[t!] 
		\centering
		\includegraphics[width= 1\linewidth]{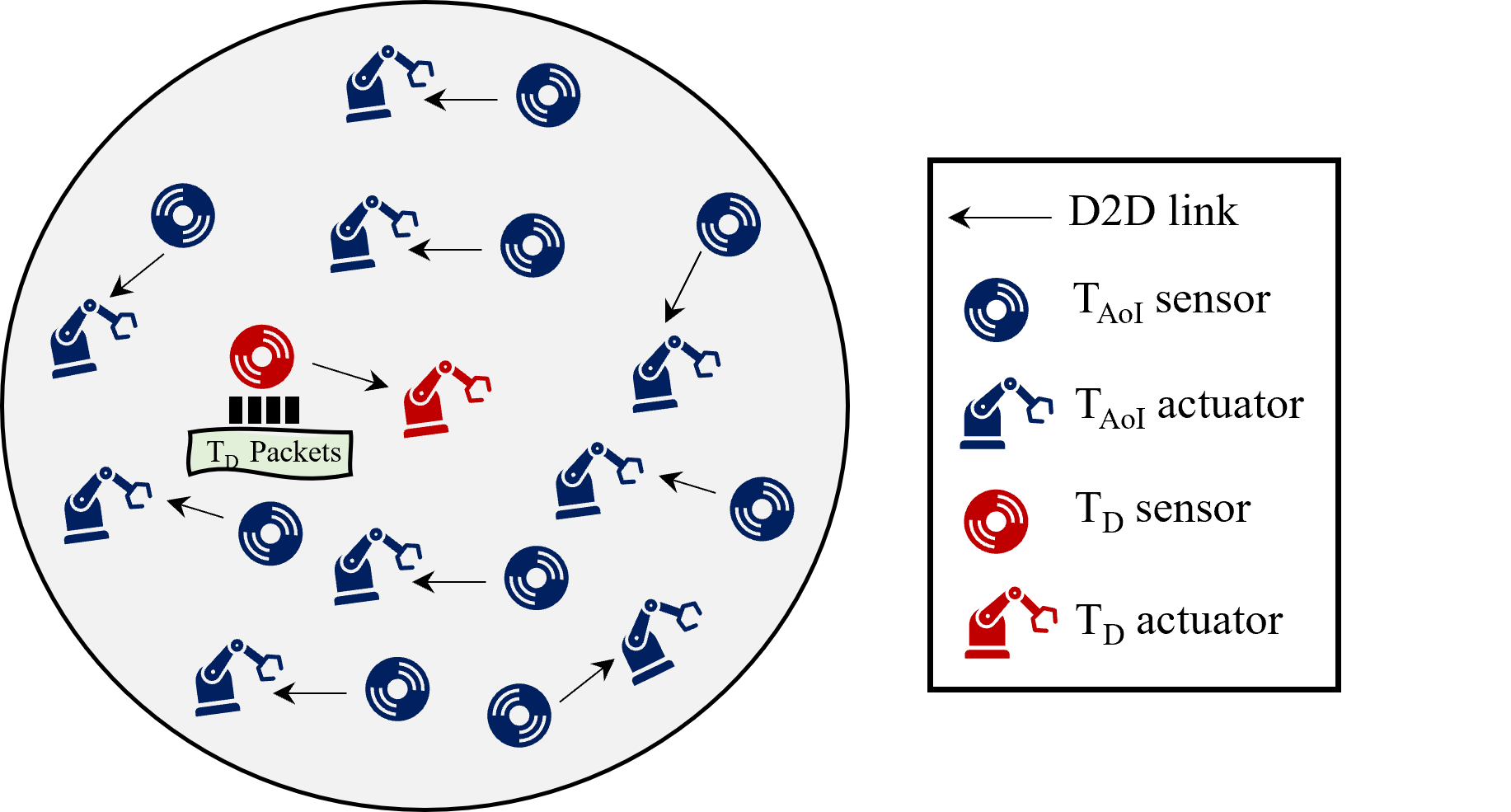}
		\caption{The network topology with multiple $\mathrm{T_{AoI}}$ pairs and a single $\mathrm{T_{D}}$ pair distributed in a circular region $\mathcal{C}$ with a radius $R$. \label{network}}
	\end{figure}

\subsection{Network Topology and Communication Model}

We consider a heterogeneous IIoT network comprising a set of sensor-actuators (D2D) pairs that are randomly distributed in a circular region $\mathcal{C}$ with a radius $R$ as depicted in Fig.~\ref{network}.
The network supports two traffic flows, namely AoI-sensitive flow ($\mathrm{T_{AoI}}$) and delay-sensitive flow ($\mathrm{T_{D}}$).
The $\mathrm{T_{AoI}}$ traffic represent the status updates of an underlying time-varying process, where the $\mathrm{T_{D}}$ traffic denotes event reporting about a certain incident. 

In our model, we consider a single D2D pair for transmitting the $\mathrm{T_{D}}$ traffic which characterized by a strict delay performance.
The $\mathrm{T_{D}}$ transmitter is fixed in predefined location where the $\mathrm{T_{D}}$ receiver is located at the origin of considered network region with a distance $d_D$.
The $\mathrm{T_{D}}$ transmitter has an infinite output queue where the arrival process of the $\mathrm{T_{D}}$ packets follows a Poisson process with average rate $\lambda_D$ packets/slot, which captures the scenario where such traffic can be triggered by the occurrence of some random incident~\cite{poission}.
The $\mathrm{T_{D}}$ packets are transmitted in First-Come-First-Served (FCFS) basis, and a packet is removed from the buffer if it is successfully transmitted, which is acknowledged by the $\mathrm{T_{D}}$ receiver through an ACK feedback. If no ACK is received within a predefined timeout, the packet is retransmitted. We assume that the ACK transmission is instantaneous and error-free, which is common assumption in the literature~\cite{Ack_free}.

Further, we consider a set of $\mathrm{T_{AoI}}$ transmitters randomly scattered according to a homogeneous Poisson Point Process (PPP) $\phi_A$ with spatial density $\lambda_A$, with $\phi_A=\{x_i=\mathbb{R}^2, \forall i\in \mathbb{N^+}, i\geq 1\}$ \cite{PPP}, where $x_i$ is the location of the $i$-th $\mathrm{T_{AoI}}$ transmitter. 
Each $\mathrm{T_{AoI}}$ transmitter has a dedicated receiver which is located at a distance $d_A$ in a random orientation. The $\mathrm{T_{AoI}}$ nodes report status updates following the generate-at will model~\cite{AoI_def, arrival}, i.e., the $\mathrm{T_{AoI}}$ transmitter is activated with a certain probability, and once becoming active, it samples a fresh packet and transmits it to the $\mathrm{T_{AoI}}$ receiver\footnote{The generate-at-will model is commonly used for optimal AoI performance as it avoids unnecessary staleness to the sampled information due to queuing \cite{AoI_def, arrival, m1, m2}. However, our analysis can be extended to other arrival models.}. After a transmission attempt, the $\mathrm{T_{AoI}}$ discards the packet regardless it is a successful or failed attempt\footnote{This scenario is based on the fact that retransmitting a stale packet might be useless for the receiver. In that sense, it would be better to transmit a fresh status update about the underlying industrial process instead of consuming the communication resources for retransmitting outdated information. Adopting retransmission-aided approach is out of the scope of this paper and left as a future work.}.

The network operates in a time-slotted fashion where the generation and the transmission of a packet is aligned with the boundary of a timeslot and a transmission attempt takes a constant time, which is assumed to be equal to the duration of one timeslot.
The $\mathrm{T_D}$ transmitter freely access the channel as long as there are packets in its queue.
The communication between each D2D pair is performed via a shared communication channel in a slotted ALOHA-like access using a slot-access probability that is subject to optimization.In other words, in every slot there is a random number of interfering users that are active, comprising the $\mathrm{T_D}$ and/or the activated $\mathrm{T_{AoI}}$ transmitters. 
The collisions among simultaneously active users are not necessarily destructive, due to the capture effect~\cite{capture}.
In that sense, a packet is decoded successfully at the end of a timeslot when the Signal-to-Interference plus-Noise ratio (SINR) at the corresponding receiver exceeds the capture ratio $\beta$. The received SINR at an arbitrary receiving node $i$ is given as
\begin{equation}\label{SINR}
\text{SINR}_{i}= \frac{P_i|h_{i,i}|^{2}d_{i,i}^{-\alpha}}{\sigma^2+\sum_{j\in \mathcal{K}\setminus\{i\}}P_j|h_{j,i}|^{2}d_{j,i}^{-\alpha}},  
\end{equation}
where $P_i$ is the transmitting power of node $i$, $h_{j,i}$ represents the Raleigh fading of the channel between the transmitter $j$ and the receiver $i$ with $h_{j,i}\sim\mathcal{CN} (0,1)$, $d_{j,i}$ is the distance between the transmitter $j$ and the receiver $i$,  $\alpha$ is the path loss exponent, $\mathcal{K}$ is the set of transmitting nodes at the same timeslot and $\sigma^2$ denotes additive white Gaussian noise (AWGN) power.

    \begin{figure}[t!] 
		\centering
		\includegraphics[width= 0.9\linewidth]{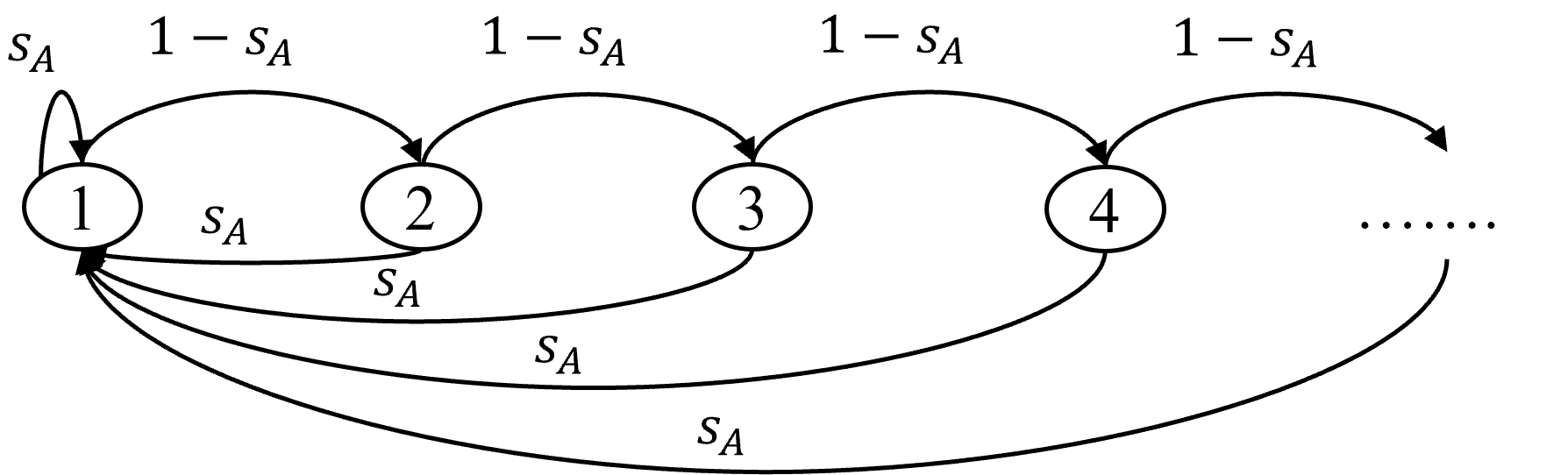}
		\caption{The DTMC model representing the evaluation of the AoI of the $\mathrm{T_{AoI}}$ traffic.  \label{DTMC-AoI}}
	\end{figure}

\subsection{Analysis of the Average AoI and the average Delay}

In this section, we analyze the Average AoI of the $\mathrm{T_{AoI}}$ traffic considering that the status updates from  all $\mathrm{T_{AoI}}$ nodes are equally important, and without loss of generality, we evaluate the average AoI at an arbitrary $\mathrm{T_{AoI}}$ receiver in discrete time.
In our model, the AoI represents the number of time slots elapsed since the last received $\mathrm{T_{AoI}}$ packet was generated. If $\Delta (t)$ denotes the AoI at the end of time slot $t$, then we have  
\begin{equation}
\Delta (t+1)=\begin{cases}
\Delta (t)+1 & \text{unsuccessful transmission}\\
1 & \text{successful transmission}.
\end{cases}
\end{equation}

Fig.~\ref{DTMC-AoI} shows the Discrete-Time Markov Chain (DTMC) model of the AoI corresponds to an arbitrary $\mathrm{T_{AoI}}$ node where each state represents the value of the AoI at the tagged $\mathrm{T_{AoI}}$ receiver. The DTMC transits from any state $j$ to 1 only upon a successful reception of a $\mathrm{T_{AoI}}$ packet, otherwise it transits to state $j+1$. Let $X_t$ represent the value of $\Delta (t)$ at time slot $t$, then the transition probability from state $j$ to state $k$ is $P_{jk}=\mathrm{Pr}(X_{t+1}=k\,\,|X_{t}=j)$, and the transition matrix $\mathbf{P_{AoI}}$ is written as
\begin{equation}
\mathbf{P_{AoI}}=
\begin{bmatrix}
s_A & 1-s_A & 0 & 0 & \dots\\
s_A & 0 & 1- s_A & 0 & \dots\\
\vdots & \vdots & \vdots & \ddots & \ddots
\end{bmatrix}, 
\end{equation}
where $s_A$ denotes  the average successful update probability per timeslot, which is defined as the probability that a $\mathrm{T_{AoI}}$ is delivered successfully to its corresponding receiver node.
The row vector $\boldsymbol{\pi_{AoI}}=[\Grave{\pi}_1, \Grave{\pi}_2, ......, \Grave{\pi}_{l-1}, \Grave{\pi}_l, ...]$ represents the steady-state probability vector of the DTMC in Fig.~\ref{DTMC-AoI}, where $\Grave{\pi}_l=\lim_{t \to +\infty}\mathds{P}(X_t)=l$ denotes the probability that the AoI is equal to $n$ at the steady state.

Using the set of equations $\boldsymbol{\pi_{AoI}}\mathbf{P_D}=\boldsymbol{\pi_{AoI}}$ and $\sum_{l}\pi_l=1$, $\pi_l$ is obtained as
\begin{equation}
\Grave{\pi_l}=s_A(1-s_A)^{(l-1)},\, \forall l.    
\end{equation}

Accordingly, the average AoI $\overline{\Delta}$ is calculated as
\begin{equation}\label{delta_AoI}
\begin{split}
\overline{\Delta}&=\sum_{l=1}^{\infty}\pi_l l=\sum_{l=1}^{\infty} l s_A(1-s_A)^{(l-1)}\\
&=\frac{s_A}{1-s_A}\sum_{l=1}^{\infty}l(1-s_A)^{l}.
\end{split}
\end{equation}

Since we have $s_A<1$, \eqref{delta_AoI} can be rewritten as 
\begin{equation}\label{AoI_FF}
  \overline{\Delta}=\frac{s_A}{1-s_A}\frac{1-s_A}{s_A^2} =\frac{1}{s_A}. 
\end{equation}

However, the average AoI cannot account for extreme AoI events occurring with very low probabilities. As mentioned in Section~\ref{intro}, the received updates of the $\mathrm{T_{AoI}}$ traffic are used for control and actuation actions, which implies certain requirements on tolerated values of AoI. In that context, we also analyze the AoI violation probability, which is the probability that the AoI exceeds a certain constraint. Denote $c$ as the target AoI constraint, then we can express the AoI violation probability as
\begin{equation}\label{AoI-violation}
\begin{split}
   &\mathrm{Pr}(\Delta>c)=1-\mathrm{Pr}(\Delta\leq c)=1-\sum_{l=1}^{c}\pi_l\\
   &=1 - \frac{s_A}{1-s_A}\sum_{l=1}^{c} (1-s_A)^l\overset{(b)}{=}(1-s_A)^c, 
\end{split}
\end{equation}
where (b) is obtained by using $\sum_{i=0}^{y}x^i = \frac{x^{y+1}-1}{x-1}$ with $x\neq 1$.

We define the average delay $D_\text{avg}$ of a $\mathrm{T_{D}}$ packet as the time elapsed from the instant the packet was generated until it is successfully delivered to its destination. Therefore, $D_\text{avg}$ consists of the queueing time of the packet $\mathrm{T_{D}}$ packet and its transmission time. According to Little's law~\cite{24}, the queuing time is calculated as $\frac{Q_\text{avg}}{\lambda_D}$, where $Q_\text{avg}$ denotes the average queue size of the $\mathrm{T_{D}}$ source. Then, the average delay $D_\text{avg}$ can be expressed as
\begin{equation}\label{delay}
D_\text{avg} = \frac{Q_\text{avg}}{\lambda_D} + \frac{1}{s_D},   
\end{equation}
where $s_D$ is the average probability of a successful transmission of a $\mathrm{T_{D}}$ packet per timeslot (average service probability), which is inversely proportional to the transmission time component of $D_\text{avg}$~\cite{tx_time}. As we can see from \eqref{AoI_FF} and \eqref{delay}, $\overline{\Delta}$ and $D_\text{avg}$ depend on the values $s_A$, $s_D$ and $Q_\text{avg}$. In the next sections, we  derive the expressions of $s_A$, $s_D$ and $Q_\text{avg}$ based on our proposed distributed backlog-aware channel access. 

\section{Analysis of the Proposed Distributed Backlog-Aware Channel Access} \label{analysis}

In the conventional slotted Aloha-based D2D communication, all nodes compete blindly and equally to transmit their data to their respective receivers via the shared channel.
For a network with dense $\mathrm{T_{AoI}}$ links (which is the scenario of our considered network model), this means that the $\mathrm{T_{D}}$ node will have a low chance to gain access to the channel.
This in turn results in a congestion problem where the buffered $\mathrm{T_{D}}$ packets suffers from an extended queuing time and a poor delay performance.
However, such performance would be unacceptable in IIoT applications where safety-critical events ($\mathrm{T_{D}}$ traffic), such as fire detection or gas leakage, must be reported with very low delay in order to perform the appropriate actuation in response.

On the other hand, low AoI is required for the $\mathrm{T_{AoI}}$ traffic in order maintain fresh knowledge of the underlying industrial process.
In that case, it is inefficient to force the $\mathrm{T_{AoI}}$ nodes to remain totally silent when the $\mathrm{T_{D}}$ node is active, in particular as the channel offers a possibility for capture effect to take place. In this regard, we propose a backlog-aware channel access to achieve a proper delay-AoI tradeoff in dense IIoT networks.
In our proposed scheme, the $\mathrm{T_{D}}$ node transmits a packet in each timeslot as long as it has a non-empty queue, while the $\mathrm{T_{AoI}}$ nodes access the channel in an opportunistic, slotted ALOHA fashion, with the slote-access probability optimized according to the status of the queue size of the $\mathrm{T_{D}}$ node. Denoting $Q$ as the random variable representing the queue size of the $\mathrm{T_{D}}$ node, the introduced access model is defined by the following cases. 

\subsection*{Case 1: Empty Queue ($Q = 0$)}

When the queue of the $\mathrm{T_{D}}$ node is empty, i.e $Q = 0$, the $\mathrm{T_{AoI}}$ nodes access the channel with a probability $p_1$. Let $\phi_{A_0}$ represent the locations of the active $\mathrm{T_{AoI}}$ transmitters when  the $\mathrm{T_{D}}$ node is silent.
Leveraging the 
thinning property of Poisson processes~\cite{thinning}, $\phi_{A_0}$ follows a homogeneous PPP with intensity $p_1 \lambda_A$ and we have $\mathcal{K} = \phi_{A_0}$.
According to Slivnyak’s theorem~\cite{thinning} and without loss of generality, the following analysis focuses on a typical active $\mathrm{T_{AoI}}$ pair.
For the considered channel model and conditioned on the spatial realization $\phi_{A_0}$, the successful decoding probability $p_{A_0}$ at the tagged $\mathrm{T_{AoI}}$ receiver when $Q = 0$ is obtained as
\begin{equation}\label{A0}
\begin{split}
p_{A_0}&=\mathrm{Pr(SINR}_{i}>\beta\,|\,\mathcal{K}=\phi_{A_0})\\
&=\mathrm{Pr}\left(\frac{P_2|h_{i,i}|^{2}d_{A}^{-\alpha}}{\sigma^2+\sum_{j\in \phi_{A_0}\setminus\{i\}}P_2|h_{j,i}|^{2}d_{j,i}^{-\alpha}}>\beta\right)\\
&\overset{(a)}{=}\mathrm{exp}\left(-\frac{\pi p_1\lambda_A d_A^2 \beta^{\frac{2}{\alpha}}}{\mathrm{sinc}(\frac{2}{\alpha})}\right) \mathrm{exp}\left(-\frac{\beta\sigma^2 d_A^\alpha}{P_2}\right),
\end{split}
\end{equation}
where $P_2$ is the transmitting power of the $\mathrm{T_{AoI}}$ nodes. The result in ($a$) follows by leveraging the Rayleigh distribution of the channel gain, i.e., $|h_{j,i}|^2\sim\mathrm{exp(1)}$ and the the probability generating functional (PGFL) of the PPP~\cite{PPP}.

\subsection*{Case 2: Moderate Backlog ($1\leq Q\leq M)$}
We denote $M$ as the backlog threshold that will determine the activity limit for the $\mathrm{T_{AoI}}$ nodes. 
When the queue is non-empty,   
the $\mathrm{T_{D}}$ node transmits in each timeslot, while the $\mathrm{T_{AoI}}$ nodes attempt to access the channel with a probability $p_2$.
We assume that $p_2 < p_1$ ,
as the $\mathrm{T_{AoI}}$ nodes should cause, as well as experience a lower interference level on the channel. 
Let $x_D$ represent the location of the $\mathrm{T_{D}}$ node and $\phi_{A_1}$ denote the locations of the active $\mathrm{T_{AoI}}$ nodes that follow a homogeneous PPP with intensity $p_2 \lambda_A$.
Then we have $\mathcal{K}=\{x_D \cup \phi_{A_1}\}$.
In this case, the successful decoding probability $p_{D_1}$ at the $\mathrm{T_{D}}$ receiver, conditioned on the spatial realization $\phi_{A_1}$, is given as
\begin{equation}\label{D1}
\begin{split}
p_{D_1}&=\mathrm{Pr(SINR}_D>\beta\,|\,\mathcal{K}=\{x_D \cup \phi_{A_1})\\
&=\mathrm{Pr}\left(\frac{P_1|h_{D,D}|^{2}d_{D}^{-\alpha}}{\sigma^2+\sum_{j\in \phi_{A_1}}P_2|h_{j,D}|^{2}d_{j,D}^{-\alpha}}>\beta\right)\\
&=\mathrm{exp}\left(-\frac{\pi p_2\lambda_A d_D^2 \left(\beta\frac{P_2}{P_1}\right)^{\frac{2}{\alpha}}}{\mathrm{sinc}(\frac{2}{\alpha})}\right) \mathrm{exp}\left(-\frac{\beta\sigma^2 d_D^\alpha}{P_1}\right).
\end{split}
\end{equation}

For the active $\mathrm{T_{AoI}}$ nodes, we derive the successful decoding probability at an arbitrary $\mathrm{T_{AoI}}$ receiver using the following proposition.

\textit{Proposition 1: When $1\leq Q\leq M$, the successful decoding probability $p_{A_1}$ at an arbitrary $\mathrm{T_{AoI}}$ node is given by}
\begin{equation}\label{A1}
p_{A_1}= \mathrm{exp} \left(-\frac{\pi p_2 \lambda_A d_A^2 \beta^\frac{2}{\alpha}}{\mathrm{sinc}(\frac{2}{\alpha})}\right)
 \frac{\mathrm{exp}\left(-\frac{\beta \sigma^2d_A^2}{P_2}\right)}{1+\frac{d_A^2}{\mathbb{E}[d_{D,i}]^2}\left(\beta\frac{P_1}{P_2}\right)^\frac{2}{\alpha}}
\end{equation}
\textit{where we have} 
\begin{equation*}
\E\left[d_{D,i}\right]=\int_0^{2\pi} \frac{1}{2 \pi} \int_0^R \frac{2r}{R^2} \sqrt{r^2 + d_{D,0}^2-2d_0d_{D,0} \cos{\theta}} \mathrm{d}r \mathrm{d}\theta.  
\end{equation*}

\textit{Proof}: See Appendix A.

\subsection*{Case 3: Congested Queue ($Q>M$)}

When the queue size of the $\mathrm{T_{D}}$ node exceeds the backlog threshold $M$, i.e., $Q>M$, then all the $\mathrm{T_{AoI}}$ nodes remain silent and only the $\mathrm{T_{D}}$ node transmits.
In this case, the successful decoding probability $p_{D_0}$ at the $\mathrm{T_{D}}$ receiver is given by 
\begin{equation}\label{D0}
\begin{split}
p_{D_0}&=\mathrm{Pr(SINR}_D>\beta\,|\,\mathcal{K}=x_D)=\mathrm{Pr}\left(\frac{P_1|h_{D,D}|^2d_D^{-\alpha}}{\sigma^2}>\beta\right)\\
&=\mathrm{exp}\left(-\frac{\beta\sigma^2 d_D^\alpha}{P_1}\right).
\end{split}
\end{equation}
Note that $p_{D_0}>p_{D_1}$.
Exploiting \eqref{A0}, \eqref{D1}, \eqref{A1}, and \eqref{D0}, we can express $s_A$ and $s_D$ as follows
\begin{equation}\label{success_AA}
\begin{split}
 s_A&= p_1\mathrm{Pr(SINR}_{i\in \phi_{A_0}}>\beta\,|\,Q=0)\, \mathrm{Pr(Q=0)}\\
  &+ p_2 \mathrm{Pr(SINR}_{i\in \phi_{A_1}}>\beta\,|\,1\leq Q \leq M)\, \mathrm{Pr(1\leq Q \leq M)}\\
  &= p_1 p_{A_0}\, \mathrm{Pr(Q=0)} + p_2p_{A_1}\mathrm{Pr(1\leq Q \leq M)}.
  \end{split}
\end{equation}

\begin{equation}\label{service}
s_D = \frac{\mathrm{Pr(1\leq Q \leq M)} p_{D_1}+\mathrm{Pr(Q>M)} p_{D_0}}{\mathrm{Pr(1\leq Q \leq M)}  + \mathrm{Pr(Q>M)}}. 
\end{equation}
As indicated by \eqref{success_AA} and \eqref{service}, both $s_A$ and $s_D$ depend on the distribution of the random variable $Q$, which is analyzed in the next section.

    \begin{figure}[t!] 
		\centering
		\includegraphics[width= 1\linewidth]{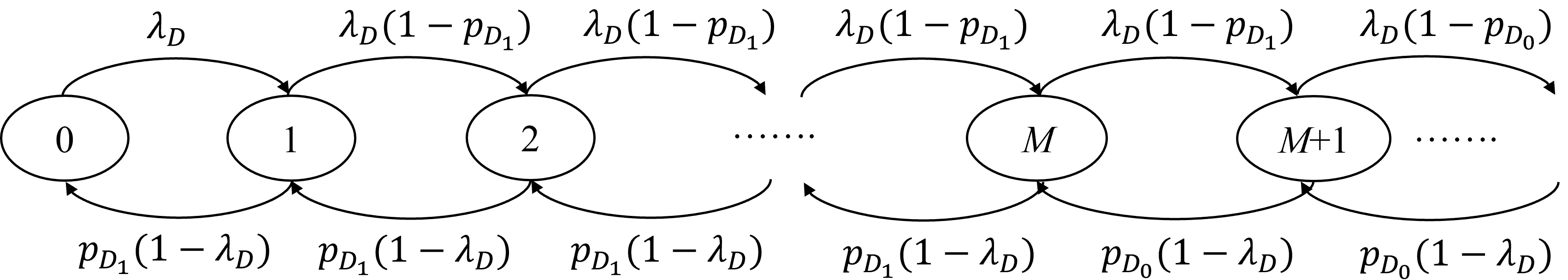}
		\caption{DTMC model of the queue at the $\mathrm{T_{D}}$ node.  \label{DTMC_queue}}
	\end{figure}

\section{The Distribution of the Queue Size $Q$}
In this section, we analyze the distribution of the queue size $Q$ to derive $\mathrm{Pr}( Q = 0)$,  $\mathrm{Pr}(1\leq Q \leq M)$, and $\mathrm{Pr}( Q > M)$. The evaluation of the queue size $Q$ can be modelled by the DTMC shown in Fig.~\ref{DTMC_queue}, where each state represents the queue size. The success probability is $p_{D_1}$  when $1\leq Q \leq M$, while it is $p_{D_0}$ when $Q > M$. 

Let $\boldsymbol{\pi_D}=\{\pi_0, \pi_1, .....\}$ represent the steady-state probability vector of the queue size of the $\mathrm{T_{D}}$ node where $\pi_n=\lim_{t \to \infty}\mathrm{Pr}(Q=n)$ is the steady-state probability of having $n$ packets at the queue. The stationary distribution of the queue size of the $\mathrm{T_{D}}$ node is given by the following lemma:

\textit{Lemma 1: According to the DTMC in Fig.~\ref{DTMC_queue}, the steady-state probability $\pi_n$ is given as}
\begin{equation}\label{Qi}
\pi_n = 
\begin{cases}
\frac{\lambda_D^n (1-p_{D_1})^{(n-1)}}{p_{D_1}^n(1-\lambda_D)^n} \pi_0, & 1\leq n \leq M\\
\frac{\lambda_D^n (1-p_{D_1})^M (1-p_{D_0})^{(n-M-1)}}{p_{D_1}^M p_{D_0}^{(n-M)}(1-\lambda_D)^n} \pi_0, &  n>M
\end{cases}
\end{equation}
\textit{where $\pi_0$ is given as}
\begin{equation}\label{Q0}
\pi_0 = 
\begin{cases}
\frac{(p_{D_1}-\lambda_D)(p_{D_0}-\lambda_D)}{p_{D_1}p_{D_0}-\lambda_D p_{D_1}-\lambda_D \left[\frac{\lambda_D(1-p_{D_1})}{(1-\lambda_D)p_{D_1}}\right]^M(p_{D_0}-p_{D_1})} & \lambda_D \neq p_{D_1}\\
\frac{p_{D_0}-p_{D_1}}{p_{D_1}+(p_{D_0}-p_{D_1})\frac{M+1-p_{D_1}}{1-p_{D_1}}} & \lambda_D = p_{D_1} 
\end{cases}
    \end{equation}

\textit{Proof:} See Appendix B.

Note that the queue of the  $\mathrm{T_{D}}$ node is stable under the condition that $\lambda_D < p_{D_0}$ (See Appendix C).

Based on the results of the steady-state probabilities in Lemma~1, the distribution of the queue size $Q$ can be given as follows:

\textit{Lemma 2: For a stable queue ($\lambda_D < p_{D_0}$), and if $\lambda_D \neq p_{D_1}$\footnote{Our derived expressions also hold for $\lambda_D = p_{D_1}$ by substituting $\pi_0$ with its corresponding expression in \eqref{Q0}.}, then the steady-state probabilities for the queue size are}
\begin{equation}\label{q1MM}
\mathrm{Pr}(1\leq Q \leq M) = \frac{\lambda_D (1-\psi^M)(p_{D_0}-\lambda_D)}{p_{D_1}p_{D_0}- \lambda_D p_{D_1}-\lambda_D \psi^M (p_{D_0}-p_{D_1})},
\end{equation}
\textit{and we also obtain}
\begin{equation}\label{qMM}
\mathrm{Pr}(Q>M) = \frac{\lambda_D \psi^M(p_{D_1}-\lambda_D)}{p_{D_1}p_{D_0}- \lambda_D p_{D_1}-\lambda_D \psi^M (p_{D_0}-p_{D_1})},
\end{equation}
\textit{where we have}
\begin{equation*}
\psi=\left(\frac{\lambda_D (1-p_{D_1})}{(1-\lambda_D) p_{D_1}}\right).   
\end{equation*}
\textit{Proof:} See Appendix D.

Based on the results obtained from Lemma~1 and Lemma~2, the average queue size $Q_\text{avg}$ can be obtained from the following theorem:

\textit{Theorem 1: The average queue size of the $\mathrm{T_{D}}$ node is given by}
\begin{equation}\label{size}
 Q_\text{avg} = \sum_{n=1}^{\infty} n \pi_n  = \frac{Q_1+Q_2}{p_{D_0}p_{D_1}-\lambda_D p_{D_1}-\lambda_D \psi^M (p_{D_0}-p_{D_1})},    
\end{equation}
\textit{where we have}
\begin{equation}
\begin{split}
Q_1 &= \lambda_D (1-\lambda_D)p_{D_1}\frac{p_{D_0-\lambda_D}}{p_{D_1}-\lambda_D}\\
&.\left(M\psi^{(M+1)}-\psi^M(M+1)+1\right),\\
Q_2 &=\psi^M \lambda_D (p_{D_1}-\lambda_D)\left(M+\frac{p_{D_0}(1-\lambda_D)}{p_{D_0}-\lambda_D}\right). \\
\end{split}
\end{equation}
\textit{Proof:} See Appendix E
\setcounter{equation}{20}
	\newcounter{storeeqcounter}
	\newcounter{tempeqcounter}
	\addtocounter{equation}{1}%
	\setcounter{storeeqcounter}%
	{\value{equation}}
	\begin{figure*}[!t]
		\normalsize
		\setcounter{tempeqcounter}{\value{equation}} 
		
		\begin{IEEEeqnarray}{rCl} 
			\setcounter{equation}{\value{storeeqcounter}} 
			\label{s_A_total}
			\begin{split}
s_A =& p_1 \mathrm{exp}\left(-\frac{\pi p_1\lambda_A d_A^2 \beta^{\frac{2}{\alpha}}}{\mathrm{sinc}(\frac{2}{\alpha})}\right) \mathrm{exp}\left(-\frac{\beta\sigma^2 d_A^\alpha}{P_2}\right) \left(\frac{(p_{D_1}-\lambda_D)(p_{D_0}-\lambda_D)}{p_{D_1}p_{D_0}-\lambda_D p_{D_1}-\lambda_D \left[\frac{\lambda_D(1-p_{D_1})}{(1-\lambda_D)p_{D_1}}\right]^M(p_{D_0}-p_{D_1})}\right)\\
+& p_2 \mathrm{exp} \left(-\frac{\pi p_2 \lambda_A d_A^2 \beta^\frac{2}{\alpha}}{\mathrm{sinc}(\frac{2}{\alpha})}\right)
 \frac{\mathrm{exp}\left(-\frac{\beta \sigma^2d_A^2}{P_2}\right)}{1+\frac{d_A^2}{\mathbb{E}[d_{D,i}]^2}\left(\beta\frac{P_1}{P_2}\right)^\frac{2}{\alpha}}  \left(\frac{\lambda_D (1-\psi^M)(p_{D_0}-\lambda_D)}{p_{D_1}p_{D_0}- \lambda_D p_{D_1}-\lambda_D \psi^M (p_{D_0}-p_{D_1})}\right).
			\end{split}
		\end{IEEEeqnarray}

		\begin{IEEEeqnarray}{rCl}
	\label{s_D_total}
			\begin{split}
s_D &= \frac{\mathrm{exp}\left(-\frac{\pi p_2\lambda_A d_D^2 \left(\beta\frac{P_2}{P_1}\right)^{\frac{2}{\alpha}}}{\mathrm{sinc}(\frac{2}{\alpha})}\right) \mathrm{exp}\left(-\frac{\beta\sigma^2 d_D^\alpha}{P_1}\right) \left(\frac{\lambda_D (1-\psi^M)(p_{D_0}-\lambda_D)}{p_{D_1}p_{D_0}- \lambda_D p_{D_1}-\lambda_D \psi^M (p_{D_0}-p_{D_1})}\right)}{1-\left(\frac{(p_{D_1}-\lambda_D)(p_{D_0}-\lambda_D)}{p_{D_1}p_{D_0}-\lambda_D p_{D_1}-\lambda_D \left[\frac{\lambda_D(1-p_{D_1})}{(1-\lambda_D)p_{D_1}}\right]^M(p_{D_0}-p_{D_1})}\right)}\\
&+\frac{\mathrm{exp}\left(-\frac{\beta\sigma^2 d_D^\alpha}{P_1}\right)\left(\frac{\lambda_D \psi^M(p_{D_1}-\lambda_D)}{p_{D_1}p_{D_0}- \lambda_D p_{D_1}-\lambda_D \psi^M (p_{D_0}-p_{D_1})}\right)}{1-\left(\frac{(p_{D_1}-\lambda_D)(p_{D_0}-\lambda_D)}{p_{D_1}p_{D_0}-\lambda_D p_{D_1}-\lambda_D \left[\frac{\lambda_D(1-p_{D_1})}{(1-\lambda_D)p_{D_1}}\right]^M(p_{D_0}-p_{D_1})}\right)}
			\end{split}
		\end{IEEEeqnarray}
		\setcounter{equation}{\value{tempeqcounter}} 
		\hrulefill
	\end{figure*}
	\setcounter{equation}{22}

Hence, based on \eqref{q1MM} and \eqref{qMM}, we can obtain $s_A$ and $s_D$ as shown in \eqref{s_A_total} and \eqref{s_D_total}. Then, by substituting $s_A$ in \eqref{AoI_FF}, we can obtain $\Delta$. Similarly, by substituting $s_D$ and $Q_{avg}$ in \eqref{delay}, we can obtain $D_{avg}$.

As we can see from \eqref{success_AA} and \eqref{AoI_FF}, when the $\mathrm{T_{D}}$ link is inactive, i.e., $Q = 0$, the age of the $\mathrm{T_{AoI}}$ nodes mainly depends on the access probability $p_1$ for a given value of the transmission power $P_2$. In that case, if we consider $\Delta$ as a function of $p_1$, then there is an optimal value $p_1^*$ that minimizes the average age $\Delta$, which is equivalent to  $p_1^* = \underset{p_1\in[0,1]}{\mathrm{arg\,max}}\,\,s_A$. So we have the following optimization problem
\begin{equation}\label{opti_p1}
\begin{split}
&\underset{p_1}{\mathrm{min}}\,\,\, p_1 \mathrm{exp}\left(-\frac{\pi p_1\lambda_A d_A^2 \beta^{\frac{2}{\alpha}}}{\mathrm{sinc}(\frac{2}{\alpha})}\right) \mathrm{exp}\left(-\frac{\beta\sigma^2 d_A^\alpha}{P_2}\right)\\
& \mathrm{subject\, to} \,\,\,\,\, 0<p_1\leq1.
\end{split}
\end{equation}
The optimization problem in \eqref{opti_p1} can be solved via the first-order optimality condition~\cite{first_order_opt}, which yields
\begin{equation}
\begin{split}
1-& p_1   \frac{\pi \lambda_A d_A^2 \beta^{\frac{2}{\alpha}}}{\mathrm{sinc}(\frac{2}{\alpha})} = 0\\
&\Rightarrow p_1 = \frac{\mathrm{sinc}(\frac{2}{\alpha})} {\pi \lambda_A d_A^2 \beta^{\frac{2}{\alpha}}}\\
&p_1^* = \mathrm{min\,\left\{\frac{\mathrm{sinc}(\frac{2}{\alpha})} {\pi \lambda_A d_A^2 \beta^{\frac{2}{\alpha}}},1\right\}}.
\end{split}
\end{equation}

By setting the obtained optimal access probability $p_1^*$ in \eqref{success_AA}, and by fixing the arrival rate $\lambda_D$ and the transmission power $P_1$, we can see that the average AoI $\Delta$ mainly depends on the access probability $p_2$ and the transmission power $P_2$. 

\textit{Remark 1: The average delay $D_{avg}$ is an independent function of $p_1$ given a certain value of $\lambda_D$. On the other hand, increasing $p_2$ and/or $P_2$ would in turn improve $\Delta$ for the $\mathrm{T_{AoI}}$ nodes but at the cost of increased delay $D_{avg}$ for the $\mathrm{T_{D}}$ node due to the increased level of interference. This is because the probability $p_{D_1}$ is decreased accordingly leading to increased queuing time at the $\mathrm{T_{D}}$ node. Thus, the average delay $D_{avg}$ is an increasing function of  $p_2$ and $P_2$.}

For a given delay constraint $D_{max}$ of the $\mathrm{T_{D}}$ traffic, there exists an optimal pair $(p_2^*, P_2^*)$ that minimizes the average AoI $\Delta$. Considering the delay constraint $D_{max}$ and the queue stability condition ($\lambda_D < p_{D_0}$) of the $\mathrm{T_{D}}$ node, the optimal pair $(p_2^*, P_2^*)$ lies within a feasible region $F_R$ that can defined as   
\begin{equation}\label{region}
  F_R = \left\{(p_2, P_2)\,:\,D_{avg}<D_{max},\, \lambda_D < p_{D_0}\right\}.  
\end{equation}

In order to find  the optimal pair $(p_2^*, P_2^*)$, we formulate the following optimization problem
\begin{equation}\label{opti_p2}
\begin{split}
\underset{(p_2,P_2)}{\mathrm{min}}&\,\,\, \Delta \\
 \mathrm{subject\, to} \,\,\,\,&\, D_{avg}(p_2, P_2)<D_{max}\\
&0<p_2\leq 1\\
&0<P_2\leq P_{max},
\end{split}
\end{equation}
where $P_{max}$ denotes the maximum transmission power of the $\mathrm{T_{AoI}}$ nodes. Deriving closed form solutions for the optimal values $(p_2^*, P_2^*)$ is a hard problem due to the complexity of the corresponding expressions of $s_D$, $s_A$ and  $D_{avg}$. Therefore, in the next section we do a numerical analysis of the solution to the optimization problem in~\eqref{opti_p2}. 

\section{Results and Discussion}
\label{sec:results}
\begin{table}[t!]
		\centering
		\caption{Evaluation parameters.}
		\label{t1}
		\begin{tabular}{ll}
			\toprule
			Parameter & Value \\
				\midrule
			$\lambda_A$ & $2\times10^{-4}$\\
			$d_D$ & \num {100} m\\
			$d_A$ & \num {50} m\\
			$R$ & \num {300} m\\
			$\alpha$ & \num {3} \\
			$\beta$ & \num {0} dB \\
			$\sigma^2$ & \num {-90} dBm\\
			$P_1$ & \num {100} mW\\
			$P_{max}$ & \num {.02} mW \\
			$D_{max}$ & \num {5} time slots \\
			\bottomrule
		\end{tabular}	
	\end{table}

In this section, we present numerical results to evaluate the performance of the proposed backlog-aware method in terms of the average delay ($D_{avg}$) and the average AoI ($\Delta$). We also introduce numerical solutions to the optimization problem \eqref{opti_p2} under different values of $\lambda_D$ and $M$. Unless stated otherwise, we adopt the evaluation parameters described in Table~\ref{t1}.
    \begin{figure}[t!] 
		\centering
		\includegraphics[width= 0.8\linewidth]{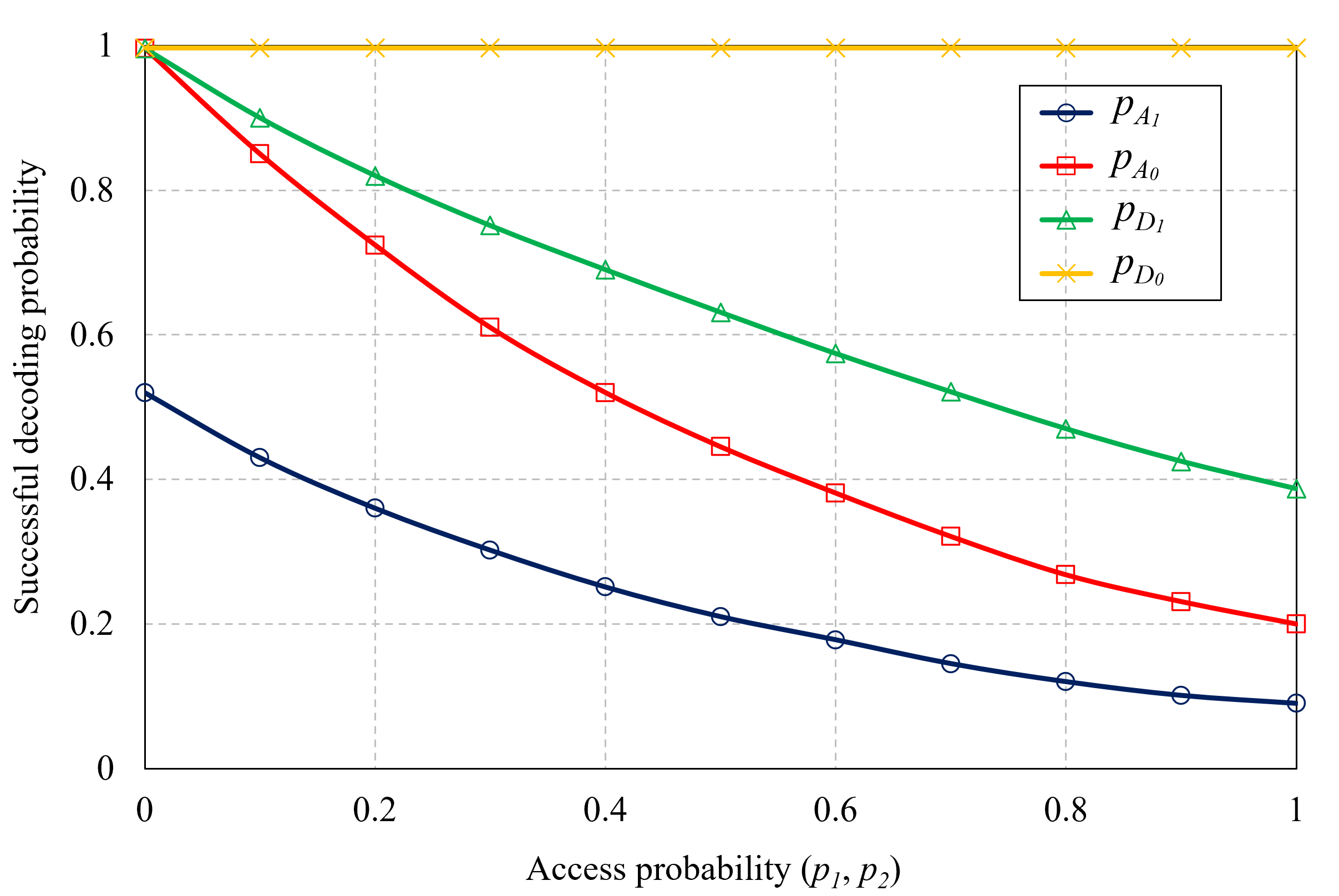}
		\caption{Evaluation of the successful decoding probabilities with varying $p_1$ and $p_2$ with $P_2 = 0.01$ mW.  \label{suc_decoding}}
	\end{figure}

    \begin{figure*}[t!] 
		\centering
		\includegraphics[width= 0.7\linewidth]{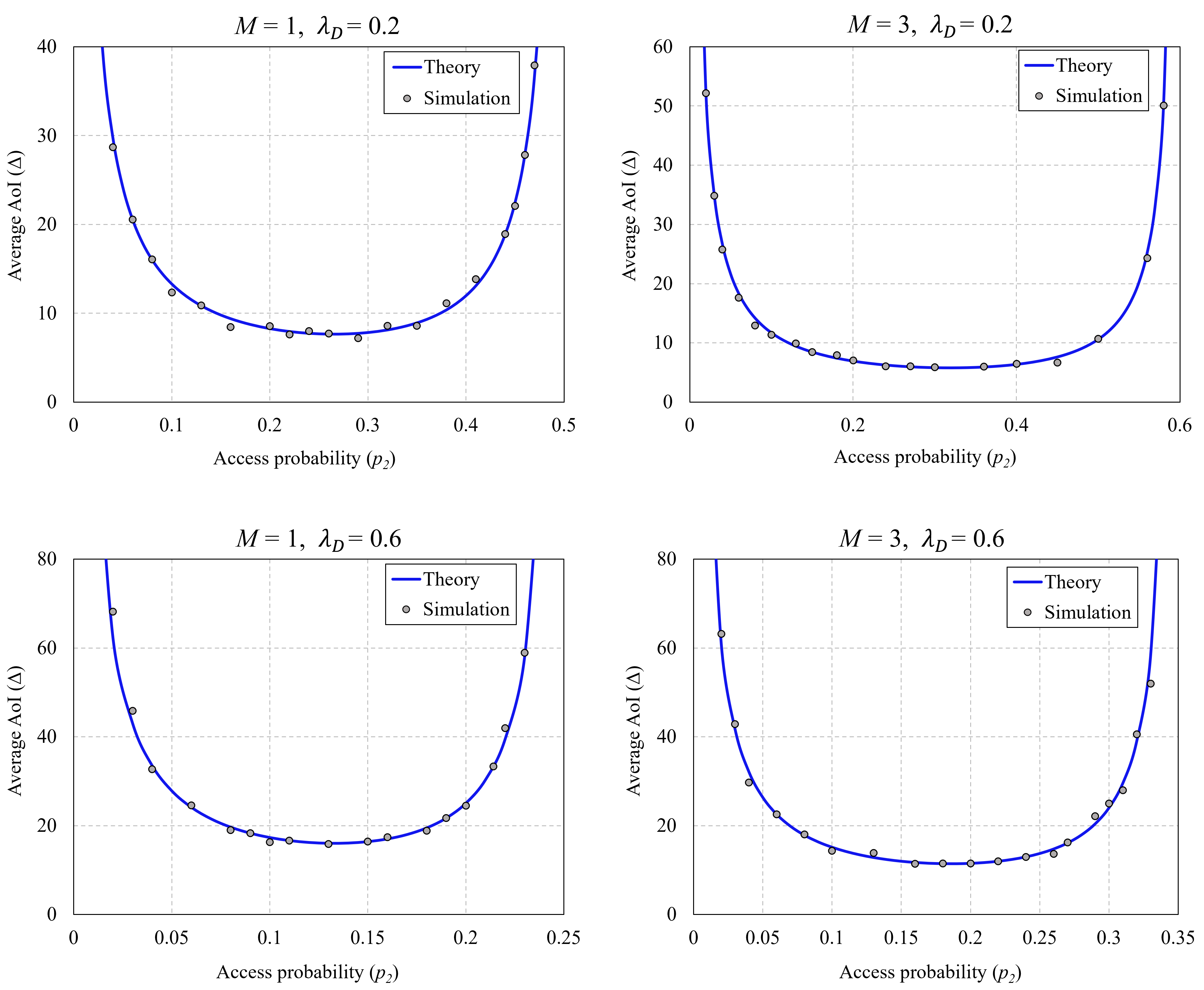}
		\caption{The average AoI performance versus $p_2$ under different values of $\lambda_D$ and $M$ with $P_2 = 0.01$ mW.  \label{AoI}}
	\end{figure*}

Fig.~\ref{suc_decoding} shows the successful decoding probabilities in \eqref{A0}, \eqref{D1}, \eqref{A1} and \eqref{D0} as a function of the access probabilities $p_1$ and $p_2$. The value of $p_{D0}$ is independent of the access probabilities $p_1$ and $p_2$, thus showing a constant value of 0.996. The successful decoding probability $p_{A0}$ depends only on $p_1$ while the values of $p_{D1}$ and $p_{A1}$ depend on $p_2$. Increasing $p_1$ implies higher collision probability between the $\mathrm{T_{AoI}}$ nodes, hence decreasing $p_{A0}$. Also, when the $\mathrm{T_{D}}$ node is active ($1\leq Q \leq M$), increasing $p_1$ would in turn decrease both $p_{A1}$ and $p_{D1}$. In order to satisfy the queue stability condition $\lambda_D < p_{D_0}$, we utilize $\lambda_D < 0.996$ for all the following results.

Fig.~\ref{AoI} presents the average AoI $\Delta$ versus the access probability $p_2$ under different values of $\lambda_D \in \{0.2, 0.6\}$ and $M \in \{1, 3\}$. Besides the numerical results obtained from \eqref{AoI_FF}, we also plot the simulation results obtained via discrete-time simulations in MATLAB to support our analytical analysis. Each simulation run lasts for $10^4$ timeslots, and we plot the average results over 100 runs. First, we can see that the simulation results match well with the numerical results, which validates our analytical analysis presented in Section~*****. Moreover, we note that $\Delta$ is not a monotonic function of $p_2$. At very low values of $p_2$, the tagged $\mathrm{T_{AoI}}$ node  is likely to be inactive for excessive period pf time. For a certain values of $\lambda_D$ and $M$, the average AoI first decreases as $p_2$ increases where the tagged $\mathrm{T_{AoI}}$ node tends to transmit updates to its corresponding actuator more frequently. However, after a certain value of $p_2$, the average AoI at the corresponding actuator starts to significantly increase due to consecutive failed transmission attempts from the tagged source as a result of increased level of contention among the $\mathrm{T_{AoI}}$ nodes. We also observe that increasing $M$ would improve the AoI as it implies a higher chance for the $\mathrm{T_{AoI}}$ source to be active. For the same $M$, higher $\lambda_D$ degrades the AoI performance as the probability $\mathrm{Pr(Q>M)}$  increases, which implies that the $\mathrm{T_{AoI}}$ nodes remain silent (\textit{Case 3} in Section~\ref{analysis}). 

    \begin{figure*}[t!] 
		\centering
		\includegraphics[width= 1\linewidth]{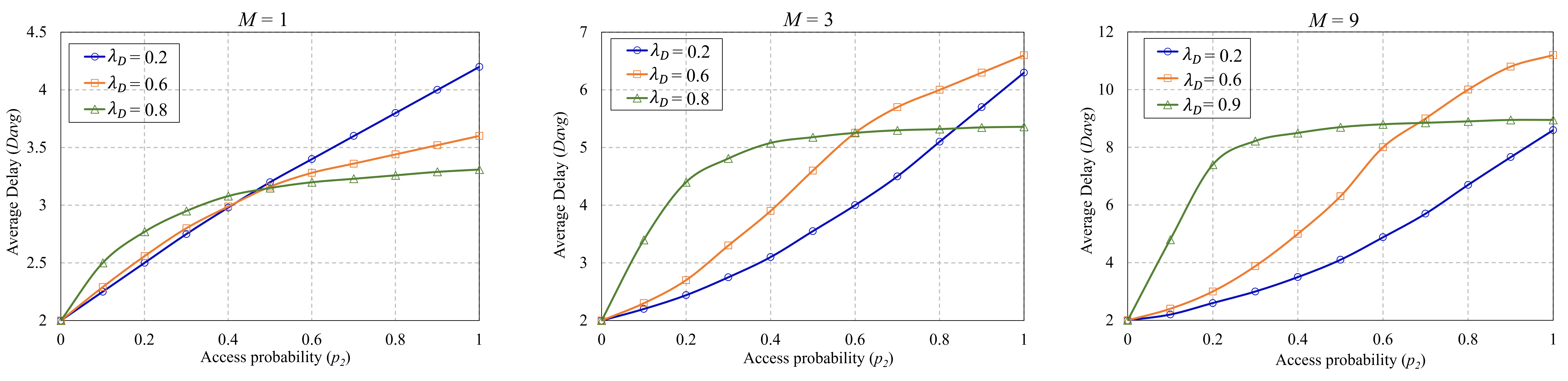}
		\caption{The average delay performance versus $p_2$ under different values of $\lambda_D$ and $M$ with $P_2 = 0.01$ mW.  \label{delay-avg}}
	\end{figure*}

Fig.~\ref{delay-avg} shows the evaluation of the average delay $D_{avg}$ against $p_2$ under different values of  $\lambda_D \in \{0.2, 0.6, 0.8\}$ and $M \in \{1, 3, 9\}$. For a certain $M$, we can observe that the average delay increases rapidly as $p_2$ increases and tends to saturate at a certain pointat high values of $\lambda_D$, which can be explained as follows. Increasing $p_2$ will in turn decrease $p_{D1}$ due to the increased interference from the $\mathrm{T_{AoI}}$ nodes, which leads to extended average queueing time at the $\mathrm{T_{D}}$ node. When $\lambda_D$ is high, it is likely that the $\mathrm{T_{AoI}}$ nodes remain inactive as the probability $\mathrm{Pr(Q>M)}$ is high, hence the $\mathrm{T_{D}}$ node would probably have interference-free transmission with a stable delay. This also explains the counter-intuitive observation in Fig.~\ref{delay-avg} where the $\mathrm{T_{D}}$ node may experience a lower delay at high value of $\lambda_D$ than the one at low value of $\lambda_D$. The figure also shows that, for the same $\lambda_D$, increasing $M$ degrades the average delay, because the backlog-aware strategy becomes weaker leading to lower $s_D$ and higher $D_{avg}$ accordingly.
    \begin{figure}[t!] 
		\centering
		\includegraphics[width= 0.9\linewidth]{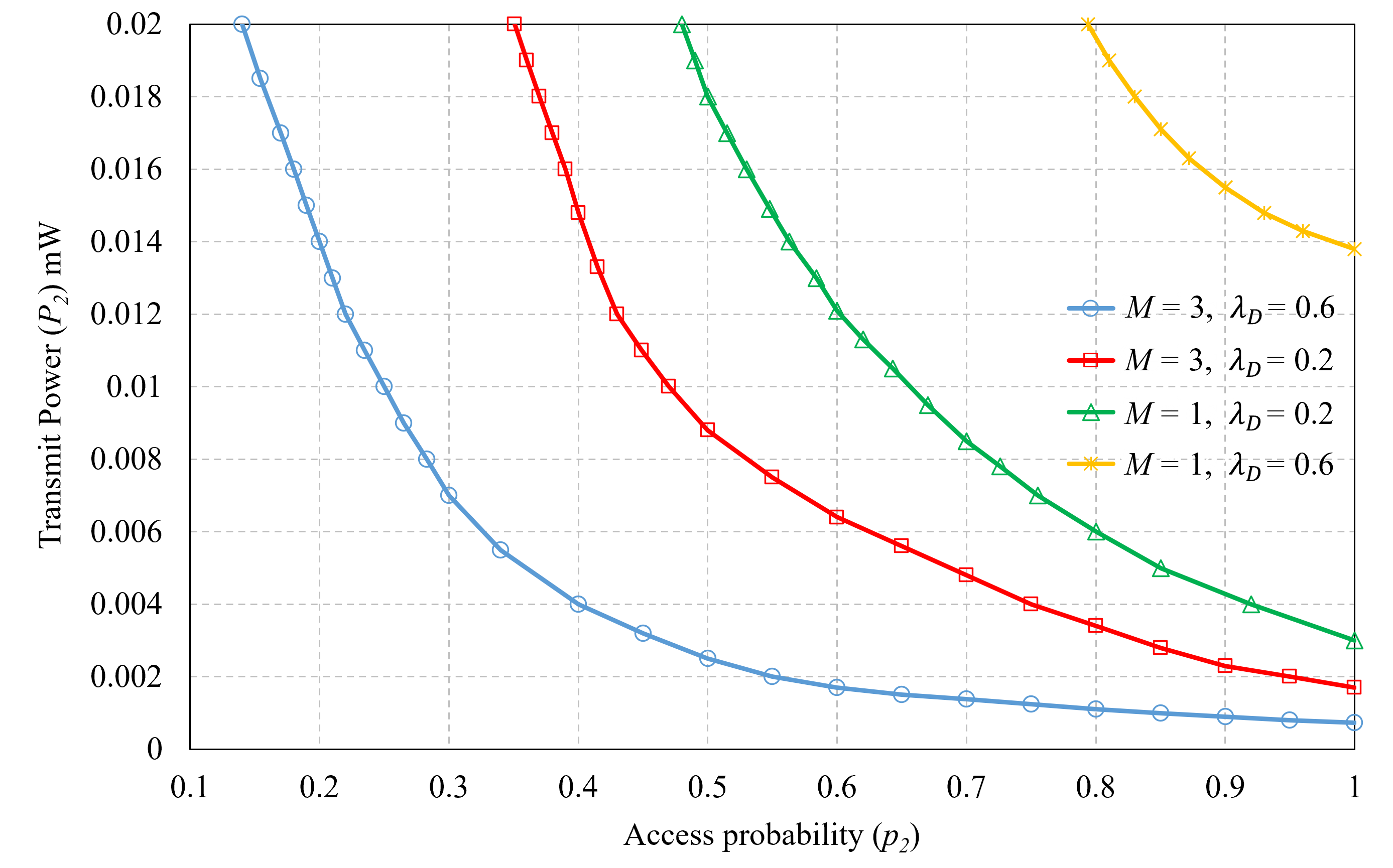}
		\caption{The feasible region $F_R$ under different values of $\lambda_D$ and $M$.  \label{feasible}}
	\end{figure}

Fig.~\ref{feasible} plots the boundary of the feasible region $F_R$ defined in \eqref{region} with $\lambda_D \in \{0.2, 0.6\}$ and $M \in \{1, 3\}$. The area below each curve comprises the possible pairs $(p_2, P_2)$ that solve the optimization problem in \eqref{opti_p2} and satisfy the delay constraint $D_{max}$. As we can see from the figure, higher $M$ implies narrower $F_R$ where the average delay becomes higher, hence narrowing down the feasible region of $(p_2, P_2)$ that satisfies the delay constraint $D_{max}$.  
\begin{table}[t]
\centering
 \caption{Numerical solutions to the optimization problem in \eqref{opti_p2}.}
 \label{optimal}
\begin{tabular}{ |l|l|l|l|l| }
\hline
$\lambda_D$ & $M$ & $p_2^*$ & $P_2^* (mW)$ & $\Delta^*$\\ \hline
\multirow{3}{*}{0.2} & 1 & 0.275 & 0.0122 & 8.22 \\ \cline{2-5}
& 3 & 0.303 & 0.0185 & 6.65 \\  \cline{2-5}
& 6 & 0.282 & 0.0161 & 6.01 \\\hline
\multirow{3}{*}{0.6} & 1 & 0.145 & 0.0091 & 13.28 \\ \cline{2-5}
& 3 & 0.192 & 0.0132 & 11.22\\  \cline{2-5}
& 6 & 0.184 & 0.0121 & 10.08 \\\hline
\multirow{3}{*}{0.8} & 1 & 0.125 & 0.0063 & 16.81 \\ \cline{2-5}
& 3 & 0.132 & 0.0072 & 15.01\\  \cline{2-5}
& 6 & 0.108 & 0.0055 & 14.09 \\\hline
\end{tabular}
\end{table}

In Table~\ref{optimal}, we present the numerical solutions ($p_2^*$, $P_2^*$) to the optimization problem in~ \eqref{opti_p2} under different values of $\lambda_D$ and $M$ and give the minimum achievable average AoI $\Delta^*$. The optimal solutions presented in this table are obtained via a 2-D numerical exhaustive search. As we can see, for the same value of $\lambda_D$, the minimum average AoI ($\Delta^*$) can be improved by increasing $M$ as the probability that the $\mathrm{T_{AoI}}$ node to be active is increased accordingly, i.e., higher $p_2^*$. However, higher $M$ increases the average delay as shown by Fig.~\ref{delay-avg}, and tighten the feasible region $F_R$ in order to satisfy the constraint $D_{max}$. For instance, when $\lambda_D = 0.2$, increasing $M$ from 1 to 3 would improve $\Delta^*$ when the $p_2^*$ is increased from 0.275 to 0.303. However, increasing $M$ further to 6 would imply decreasing $p_2^*$ to 0.282 in order to keep $p_{D1}$ sufficiently high to comply with the delay constraint of $D_{max} = 5$ timeslots. In addition, for a given value of $M$, the optimal access probability $p_2^*$ and the optimal transmission power $P_2^*$ are lower for higher $\lambda_D$ in order to reduce the interference from the $\mathrm{T_{AoI}}$ nodes and satisfy $D_{max}$.

\section{Conclusion and Future Work}
In this paper, the inherent interplay between delay and AoI is investigated for a D2D-based heterogeneous IIoT network. We proposed a distributed backlog-aware access protocol and developed closed-form expressions for the average AoI and average delay using stochastic geometry and queuing theory. We evaluated the network performance using numerical results under different parameters and obtained optimal solutions that minimize the AoI via  2-D exhaustive numerical search. As a future work, this work can be extended to adopt queue management approaches and adaptive retransmission to introduce extra enhancement to the AoI. Moreover, we consider using machine learning-based techniques to set the optimal network parameters to adapt to the dynamic industrial environment.
\label{sec:conclusions}

\section*{Acknowledgement}
This paper has received funding from the European Union’s Horizon 2020 research and innovation programme under grant agreement No. 883315.

\appendix

\subsection{Proof of Proposition 1}

    \begin{figure}[t!] 
		\centering
		\includegraphics[width= 0.5\linewidth]{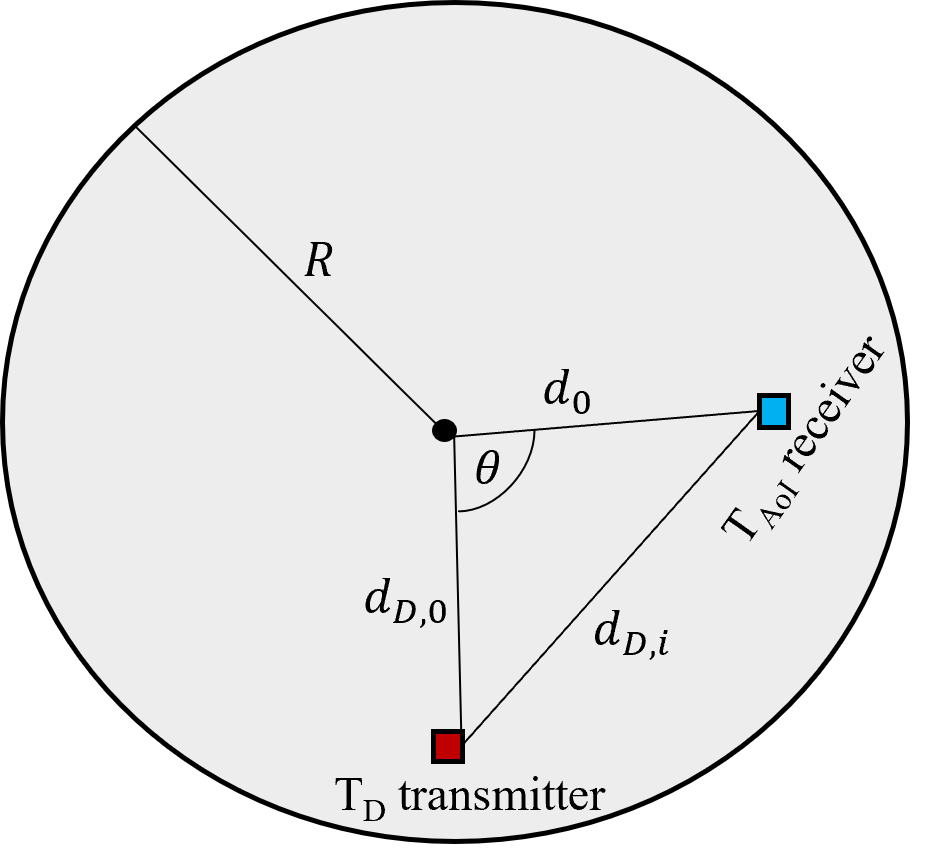}
		\caption{Geometric representation of the distribution of the $\mathrm{T_{D}}$ transmitter and an arbitrary $\mathrm{T_{AoI}}$ receiver.  \label{geometry}}
	\end{figure}

The successful update probability $p_{A_1}$ of an arbitrary $\mathrm{T_{AoI}}$ pair when the $\mathrm{T_{D}}$ node is active is given as


 \begin{equation}\label{AA}
\begin{split}
&p_{A_1}=\mathrm{Pr(SINR}_{i}>\beta\,|\,\mathcal{K}=\phi_{A_1}\cup x_D)\\
&=\mathrm{Pr}\left(\frac{P_2|h_{i,i}|^{2}d_{A}^{-\alpha}}{\sigma^2+\sum_{j\in \phi_{A_1}\setminus\{i\}}P_2|h_{j,i}|^{2}d_{j,i}^{-\alpha}+P_1|h_{D,i}|^2 d_{D,i}^{-\alpha}}>\beta\right).
\end{split}
\end{equation}
Since we have $|h_{i,i}|^2 \sim \mathrm{exp (1)}$ and $|h_{D,i}|^2 \sim \mathrm{exp (1)}$, so \eqref{AA} can be rewritten as

\begin{equation}
    \begin{split}
 p_{A_1}& = \mathrm{exp}\left(-\frac{\beta \sigma^2 d_A^\alpha}{P_2}\right) \E \left[\int_0^ \infty \mathrm{exp}\left(-\frac{P_1}{P_2}d_{D,i}^{-\alpha}x\right)e^{-x}\mathrm{d}x\right] \\
 & \cdot \underbrace{\E \left[\mathrm{exp}\left(-s\sum_{j\in \phi_{A_1}\setminus\{i\}}|h_{j,i}|^2 d_{j,i}^{-\alpha}\right)\right]}_{L_S}\\
 & = \mathrm{exp}\left(-\frac{\beta \sigma^2 d_A^\alpha}{P_2}\right) \E \left[\frac{1}{1+\frac{P_1}{P_2}\beta d_A^{\alpha}d_{D,i}^{-\alpha}}\right]\\
 & \cdot \underbrace{\E \left[\mathrm{exp}\left(-s\sum_{j\in \phi_{A_1}\setminus\{i\}}|h_{j,i}|^2 d^{-\alpha}_{j,i}\right)\right]}_{L_S},
    \end{split}
\end{equation}
 where the term $L_S$ denotes the Laplace transform of the interference component of the nodes in $\phi_{A_1}$ which can be expressed as \cite{Laplace} 

\begin{equation} \label{laplace}
\E \left[\mathrm{exp}\left(-s\sum_{j\in \phi_{A_1}\setminus\{i\}}|h_{j,i}|^2 d^{-\alpha}_{j,i}\right)\right] = \mathrm{exp}\left[-\frac{\pi p_2 \lambda_A d_A^2 \beta^{\frac{2}{\alpha}}}{\mathrm{sinc}(\frac{2}{\alpha})}\right].   
\end{equation}
Based on~\cite{approx_expec}, we can also have 
\begin{equation}\label{approx}
  \E \left[\frac{1}{1+\frac{P_1}{P_2}\beta d_A^{\alpha}d_{D,i}^{-\alpha}}\right] \simeq \frac{1}{1+\frac{d_A^2}{\E\left[d_{D,i}\right]^2} \left(\beta \frac{P_1}{P_2}\right)^{\frac{2}{\alpha}}}.  
\end{equation}
Then, with \eqref{laplace} and \eqref{approx}, $p_{A_1}$ can be expressed as

\begin{equation} \label{success_A}
    p_{A_1} \simeq \mathrm{exp}\left[-\frac{\pi p_2 \lambda_A d_A^2 \beta^{\frac{2}{\alpha}}}{\mathrm{sinc}(\frac{2}{\alpha})}\right] \frac{\mathrm{exp}\left(-\frac{\beta \sigma^2 d_A^\alpha}{P_2}\right)}{1+\frac{d_A^2}{\E\left[d_{D,i}\right]^2} \left(\beta \frac{P_1}{P_2}\right)^{\frac{2}{\alpha}}}
\end{equation}

The level of interference experienced by an arbitrary $\mathrm{T_{AoI}}$ receiver from the $\mathrm{T_{D}}$ transmitter depends on the distance $d_{D,i}$. Therefore, the term  $\E\left[d_{D,i}\right]$ is calculated by taking the expectation over all the possible locations of that $\mathrm{T_{AoI}}$ receiver within the considered region $\mathcal{C}$. Since the locations of the  $\mathrm{T_{AoI}}$ transmitters follow a homogeneous PPP model, we can consider that the locations of the $\mathrm{T_{AoI}}$ receivers approximately follow a uniform distribution in the region $\mathcal{C}$ with a radius $R$. Accordingly, the distance $d_0$ of an arbitrary $\mathrm{T_{AoI}}$ receiver from origin of $\mathcal{C}$ follow the probability distribution function $f_{d_0}(r)$ which is given by
\begin{equation}
 f_{d_0}(r) =
\begin{cases}
\frac{2r}{R^2} & \mathrm{if}\,0\leq r \leq R \\
0 & \mathrm{otherwise.}
\end{cases}
\end{equation}

If we consider the region $\mathcal{C}$ depicted in Fig.~\ref{geometry}, then we have

\begin{equation}
 d_{D,i} = \sqrt{d_0^2 + d_{D,0}^2-2d_0d_{D,0} \cos{\theta}},
\end{equation}
where $d_{D,0}$ is the distance from the $\mathrm{T_{D}}$ transmitter to the origin of the region $\mathcal{C}$, and $\theta \in \{0, 2\pi\}$ is a uniformly distributed random variable. Then, the expectation $\E\left[d_{D,i}\right]$ can be obtained as

\begin{equation} \label{expectation}
\E\left[d_{D,i}\right]=\int_0^{2\pi} \frac{1}{2 \pi} \int_0^R \frac{2r}{R^2} \sqrt{r^2 + d_{D,0}^2-2d_0d_{D,0} \cos{\theta}} \mathrm{d}r \mathrm{d}\theta.    
\end{equation}
Based on \eqref{success_A} and \eqref{expectation}, we can obtain the probability $ p_{A_1}$, which concludes the proof.

\subsection{Proof of Lemma 1}

The balance equations for the DTMC in Fig.~\ref{DTMC_queue} can be obtained as

\begin{equation}
\begin{split}
\lambda_D \pi_0 = (1-\lambda_D) p_{D_1} \pi_1 & \Leftrightarrow \pi_1 = \pi_0\frac{\lambda_D}{(1-\lambda_D)p_{D_1}}\\ 
\pi_1 \left(\lambda_D(1-p_{D_1})+(1-\lambda_D)p_{D_1}\right) &= \lambda_D \pi_0 + \pi_2 p_{D_1}(1-\lambda_D)\\
 \Leftrightarrow & \pi_2 = \pi_0 \frac{\lambda_D^2 (1-p_{D_1})}{(1-\lambda_D)^2 p_{D_1}^2}.
\end{split}
\end{equation}

Hence, for $1 \leq n \leq M$ we have

\begin{equation}
\pi_n =  \frac{\lambda_D^n (1-p_{D_1})^{(n-1)}}{p_{D_1}^n(1-\lambda_D)^n} \pi_0 ,
\end{equation}

while for $n>M$ we get

\begin{equation} \label {pi_nn}
\pi_n = \frac{\lambda_D^n (1-p_{D_1})^M (1-p_{D_0})^{(n-M-1)}}{p_{D_1}^M p_{D_0}^{(n-M)}(1-\lambda_D)^n} \pi_0
\end{equation}

Since we have $\sum_{n=0}^\infty \pi_n = 1$, with $\lambda_D \neq p_{D_1}$, the empty queue probability $\pi_0$ is expressed as

\begin{equation} \label{pi_0}
\pi_0 = \frac{(p_{D_1}-\lambda_D)(p_{D_0}-\lambda_D)}{p_{D_1}p_{D_0}-\lambda_D p_{D_1}-\lambda_D \left[\frac{\lambda_D(1-p_{D_1})}{(1-\lambda_D)p_{D_1}}\right]^M(p_{D_0}-p_{D_1})}.   
\end{equation}

When $\lambda_D = p_{D_1}$, \eqref{pi_0} is no longer valid ($\pi_0 = \frac{0}{0}$). In this case we apply L'H\^{o}spital's rule and get $\pi_0$ as

\begin{equation} \label{pi_01}
\pi_0 = \lim_{\lambda_D \to p_{D_1}} \frac{f\sp{\prime}(\lambda_D)}{h\sp{\prime}(\lambda_D)} = \frac{p_{D_0}-p_{D_1}}{p_{D_1}+(p_{D_0}-p_{D_1})\frac{M+1-p_{D_1}}{1-p_{D_1}}},
\end{equation}
where $f(\lambda_D)$ and $h(\lambda_D)$ represent the nominator and denominator of \eqref{pi_0}, respectively. Combining \eqref{pi_0} and \eqref{pi_01}, we get \eqref{Q0} which concludes the proof of \textit{Lemma 1}.

\subsection{Stability Condition of the $\mathrm{T_{AoI}}$ Node Queue}

Our basic definition for the queue stability is based on Loynes’ theorem~\cite{Loynes}. The theorem states that, a queue is stable if the average service probability is greater than average arrival probability with the condition that the arrival process and the service process are strictly jointly stationary and. Hence, in order for the queue of the $\mathrm{T_{D}}$ to be stable, we need to prove that the DTMC in Fig.~\ref{DTMC_queue}  irreducible Markov chain. Then, the average $\lambda_D$ in \eqref{pi_0} must satisfy the condition $0<\pi_0<1$. We consider the following cases of the values of $\lambda_D$ relative to the values of $p_{D_1}$ and $p_{D_0}$:
\begin{itemize}

    \item When $\lambda_D < p_{D_1} \Rightarrow \frac{\lambda_D(1-p_{D_1})}{(1-\lambda_D)p_{D_1}}<1$, then $h(\lambda_D)>p_{D_0}(p_{D_1}-\lambda_D)$. Hence, we have $\pi_0<\frac{p_{D_0}-\lambda_D}{p_{D_0}} = 1-\frac{\lambda_D}{p_{D_0}}<1$. 
    
    \item When $\lambda_D = p_{D_1}$, based on \eqref{pi_01}, it is obvious that $0<\pi_0<1$
    
    \item When $p_{D_1} < \lambda_D < p_{D_0} \Rightarrow \frac{\lambda_D(1-p_{D_1})}{(1-\lambda_D)p_{D_1}}>1$, then we have $f(\lambda_D)<0$. Also, as from the following inequality
    \begin{equation*}
\left(\frac{1-\lambda_D}{1-p_{D_1}}\right)^M \frac{p_{D_0}-\lambda_D}{p_{D_0}-p_{D_1}}<1<\left(\frac{\lambda_D}{p_{D_1}}\right)^{M+1} 
    \end{equation*}
    \begin{equation*}
     \begin{split}
&\Rightarrow p_{D_1} (p_{D_0}-\lambda_D)<\lambda_D \left(\frac{\lambda_D(1-p_{D_1})}{(1-\lambda_D)p_{D_1}}\right)^M (p_{D_0}-p_{D_1})  \\
&\Rightarrow h(\lambda_D)<0.
 \end{split}
    \end{equation*}
Thus, we have $\pi_0 = \frac{f(\lambda_D)}{h(\lambda_D)}>0$. Since  $\frac{\lambda_D(1-p_{D_1})}{(1-\lambda_D)p_{D_1}}>1$, we also have $h(\lambda_D)<p_{D_0}(p_{D_1}-\lambda_D)<0$, so we obtain $\pi_0<1-\frac{\lambda_D}{p_{D_0}}<1$   
\end{itemize}

From the discussion of the above cases, we can see that the condition $0<\pi_0<1$ is verified, and we can conclude that  the queue of $\mathrm{T_{D}}$ node is stable if $\lambda_D<p_{D_0}$.

\subsection{Proof of Lemma 2}

The probability $\mathrm{Pr}(1\leq Q \leq M)$ is obtained as

\begin{equation}
\mathrm{Pr}(1\leq Q \leq M) = \sum_{n=1}^M \pi_n.
\end{equation}

Based on \eqref{Qi} and when $\lambda_D < p_{D_0}$ and $\psi\neq1$, we get

\begin{equation}
\begin{split}
 \mathrm{Pr}(1\leq Q \leq M)& = \frac{\pi_0}{1-p_{D_1}} \sum_{n=1}^M \left(\frac{\lambda_D(1-p_{D_1})}{(1-\lambda_D)p_{D_1}}\right)^n  \\
 & = \frac{\pi_0}{1-p_{D_1}} \frac{\frac{\lambda_D(1-p_{D_1})}{(1-\lambda_D)p_{D_1}}-\left(\frac{\lambda_D(1-p_{D_1})}{(1-\lambda_D)p_{D_1}}\right)^{n+1}}{1-\frac{\lambda_D(1-p_{D_1})}{(1-\lambda_D)p_{D_1}}}\\
 & = \frac{\pi_0 \lambda_D (1-\psi^M)}{p_{D_1}-\lambda_D}\\
 & = \frac{\lambda_D (1-\psi^M)(p_{D_0}-\lambda_D)}{p_{D_1}p_{D_0}- \lambda_D p_{D_1}-\lambda_D \psi^M (p_{D_0}-p_{D_1})}.
 \end{split}
\end{equation}

For $n>M$, we obtain $\mathrm{Pr}(Q > M)$ as

\begin{equation} \label{QgeqM}
\begin{split}
   \mathrm{Pr}(Q > M)& = \sum_{n=M+1}^\infty \pi_n  = 1-\pi_0-\sum_{n=1}^M \pi_n\\  
    & = \frac{\lambda_D \psi^M(p_{D_1}-\lambda_D)}{p_{D_1}p_{D_0}- \lambda_D p_{D_1}-\lambda_D \psi^M (p_{D_0}-p_{D_1})}.
\end{split}    
\end{equation}

\subsection{Proof of Theorem 1}

The average queue size $Q_{avg}$ of the $\mathrm{T_{AoI}}$ node can be given as

\begin{equation} \label{avg_Q}
\begin{split}
 Q_{avg} &=  \sum_{n=1}^\infty n \pi_n =  \sum_{n=1}^M n \pi_n + \sum_{n=M+1}^{\infty} n \pi_n  \\  
 & = \sum_{n=1}^M n \pi_n + \sum_{n=1}^{\infty} (M+n)  \pi_{M+n} \\
 & = \underbrace{\sum_{n=1}^M n \pi_n}_{R_1} + \underbrace{M \sum_{n=1}^{\infty} \pi_{M+n}}_{R_2}+  \underbrace{\sum_{n=1}^{\infty} n \pi_{M+n}}_{R_3} 
\end{split}
\end{equation}
With $\lambda_D < p_{D_0}$ and $\psi\neq1$, the first term $R_1$ can be obtained as

\begin{equation} \label{R_1}
    \begin{split}
 \sum_{n=1}^M n \pi_n & =\sum_{n=1}^M n \frac{\lambda_D^n (1-p_{D_1})^{(n-1)}}{p_{D_1}^n(1-\lambda_D)^n} \pi_0 \\      
 & = \frac{\pi_0 \lambda_D}{(1-\lambda_D)p_{D_1}}\sum_{n=1}^M n\left[\frac{\lambda_D (1-p_{D_1})}{p_{D_1}(1-\lambda_D)}\right]^{(n-1)}\\
 & = \frac{\pi_0 \lambda_D}{(1-\lambda_D)p_{D_1}}{\left(\sum_{n=1}^M \left[\frac{\lambda_D (1-p_{D_1})}{p_{D_1}(1-\lambda_D)}\right]^{n}\right)}\sp{\prime}\\
 & = \frac{\pi_0 \lambda_D}{(1-\lambda_D)p_{D_1}} \frac{M \psi^{(M+1)}-\psi^M (M+1)+1}{\left(1-\frac{\lambda_D (1-p_{D_1})}{p_{D_1}(1-\lambda_D)}\right)^2} \\
 & = \frac{\lambda_D (1-\lambda_D)p_{D_1} \frac{p_{D_0}-\lambda_D}{p_{D_1}-\lambda_D} \left(M \psi^{(M+1)}-\psi^M (M+1)+1\right)}{p_{D_1}p_{D_0}- \lambda_D p_{D_1}-\lambda_D \psi^M (p_{D_0}-p_{D_1})}
    \end{split}
\end{equation}

Based on the results of \eqref{qMM}, the term $R_2$ is obtained as

\begin{equation} \label{R_2}
\begin{split}
M \sum_{n=1}^{\infty} \pi_{M+n} &=  M\,\mathrm{Pr}(Q > M) \\
& = \frac{\lambda_D \psi^M(p_{D_1}-\lambda_D)}{p_{D_1}p_{D_0}- \lambda_D p_{D_1}-\lambda_D \psi^M (p_{D_0}-p_{D_1})}
\end{split}
\end{equation}

Finally, the term $R_3$ can be obtained using the results from \eqref{pi_nn}

\begin{equation} \label{R_3}
\begin{split}
 \sum_{n=1}^{\infty} n \pi_{M+n} &= \frac{\pi_0 \lambda_D\psi^M}{ p_{D_0}(1-\lambda_D)} \sum_{n=1}^{\infty} n \left[\frac{\lambda_D (1-p_{D_0})}{p_{D_0}(1-\lambda_D)}\right]^{(n-1)}\\
 & = \frac{\pi_0 \lambda_D\psi^M}{ p_{D_0}(1-\lambda_D)} \frac{1}{\left(1-\frac{\lambda_D (1-p_{D_0})}{p_{D_0}(1-\lambda_D)}\right)^2}\\
 & = \frac{\lambda_D (1-\lambda_D)p_{D_0}\psi^M \frac{p_{D_1}-\lambda_D}{p_{D_0}-\lambda_D}}{p_{D_1}p_{D_0}- \lambda_D p_{D_1}-\lambda_D \psi^M (p_{D_0}-p_{D_1})}
\end{split}
\end{equation}

Substituting \eqref{R_1}, \eqref{R_2} and \eqref{R_3} in \eqref{avg_Q}, we get
\begin{equation}
 Q_{avg}=\frac{Q_1+Q_2}{p_{D_0}p_{D_1}-\lambda_D p_{D_1}-\lambda_D \psi^M (p_{D_0}-p_{D_1})},   
\end{equation}
where we have
\begin{equation}
\begin{split}
Q_1 &= \lambda_D (1-\lambda_D)p_{D_1}\frac{p_{D_0-\lambda_D}}{p_{D_1}-\lambda_D}\\
&.\left(M\psi^{(M+1)}-\psi^M(M+1)+1\right),\\
Q_2 &=\psi^M \lambda_D (p_{D_1}-\lambda_D)\left(M+\frac{p_{D_0}(1-\lambda_D)}{p_{D_0}-\lambda_D}\right). \\
\end{split}
\end{equation}

\bibliographystyle{IEEEtran}
\bibliography{main}

\begin{thebibliography}{10}
\providecommand{\url}[1]{#1}
\csname url@samestyle\endcsname
\providecommand{\newblock}{\relax}
\providecommand{\bibinfo}[2]{#2}
\providecommand{\BIBentrySTDinterwordspacing}{\spaceskip=0pt\relax}
\providecommand{\BIBentryALTinterwordstretchfactor}{4}
\providecommand{\BIBentryALTinterwordspacing}{\spaceskip=\fontdimen2\font plus
\BIBentryALTinterwordstretchfactor\fontdimen3\font minus
  \fontdimen4\font\relax}
\providecommand{\BIBforeignlanguage}[2]{{%
\expandafter\ifx\csname l@#1\endcsname\relax
\typeout{** WARNING: IEEEtran.bst: No hyphenation pattern has been}%
\typeout{** loaded for the language `#1'. Using the pattern for}%
\typeout{** the default language instead.}%
\else
\language=\csname l@#1\endcsname
\fi
#2}}
\providecommand{\BIBdecl}{\relax}
\BIBdecl

\bibitem{iot}
J.~Lin, W.~Yu, N.~Zhang, X.~Yang, H.~Zhang, and W.~Zhao, ``A survey on internet
  of things: Architecture, enabling technologies, security and privacy, and
  applications,'' \emph{IEEE Internet of Things Journal}, vol.~4, no.~5, pp.
  1125--1142, 2017.

\bibitem{iot_number}
\BIBentryALTinterwordspacing
G.~Intelligence. (2021) The mobile economy 2021. [Online]. Available:
  \url{https://www.gsma.com/mobileeconomy/}
\BIBentrySTDinterwordspacing

\bibitem{IIoT}
E.~Sisinni, A.~Saifullah, S.~Han, U.~Jennehag, and M.~Gidlund, ``Industrial
  internet of things: Challenges, opportunities, and directions,'' \emph{IEEE
  Transactions on Industrial Informatics}, vol.~14, no.~11, pp. 4724--4734,
  2018.

\bibitem{dense_IIoT}
F.~Ademaj, M.~Rzymowski, H.-P. Bernhard, K.~Nyka, and L.~Kulas, ``Relay-aided
  wireless sensor network discovery algorithm for dense industrial iot
  utilizing espar antennas,'' \emph{IEEE Internet of Things Journal}, vol.~8,
  no.~22, pp. 16\,653--16\,665, 2021.

\bibitem{D2D_IIoT}
L.~Liu and W.~Yu, ``A d2d-based protocol for ultra-reliable wireless
  communications for industrial automation,'' \emph{IEEE Transactions on
  Wireless Communications}, vol.~17, no.~8, pp. 5045--5058, 2018.

\bibitem{Hossam}
H.~Farag, E.~Sisinni, M.~Gidlund, and P.~Österberg, ``Priority-aware wireless
  fieldbus protocol for mixed-criticality industrial wireless sensor
  networks,'' \emph{IEEE Sensors Journal}, vol.~19, no.~7, pp. 2767--2780,
  2019.

\bibitem{AoI_D2D_IIoT}
M.~Li, C.~Chen, C.~Hua, and X.~Guan, ``Learning-based autonomous scheduling for
  aoi-aware industrial wireless networks,'' \emph{IEEE Internet of Things
  Journal}, vol.~7, no.~9, pp. 9175--9188, 2020.

\bibitem{oil_refinery}
P.~Gil, A.~Santos, and A.~Cardoso, ``Dealing with outliers in wireless sensor
  networks: An oil refinery application,'' \emph{IEEE Transactions on Control
  Systems Technology}, vol.~22, no.~4, pp. 1589--1596, 2014.

\bibitem{AoI_def}
Y.~Sun, E.~Uysal-Biyikoglu, R.~D. Yates, C.~E. Koksal, and N.~B. Shroff,
  ``Update or wait: How to keep your data fresh,'' \emph{IEEE Transactions on
  Information Theory}, vol.~63, no.~11, pp. 7492--7508, 2017.

\bibitem{D2D_Energy}
R.~Li, P.~Hong, K.~Xue, M.~Zhang, and T.~Yang, ``Resource allocation for uplink
  noma-based d2d communication in energy harvesting scenario: A two-stage game
  approach,'' \emph{IEEE Transactions on Wireless Communications}, vol.~21,
  no.~2, pp. 976--990, 2022.

\bibitem{Q2}
H.~H. Yang, C.~Xu, X.~Wang, D.~Feng, and T.~Q.~S. Quek, ``Understanding age of
  information in large-scale wireless networks,'' \emph{IEEE Transactions on
  Wireless Communications}, vol.~20, no.~5, pp. 3196--3210, 2021.

\bibitem{Q3}
R.~D. Yates, ``The age of information in networks: Moments, distributions, and
  sampling,'' \emph{IEEE Transactions on Information Theory}, vol.~66, no.~9,
  pp. 5712--5728, 2020.

\bibitem{q_management}
M.~Costa, M.~Codreanu, and A.~Ephremides, ``On the age of information in status
  update systems with packet management,'' \emph{IEEE Transactions on
  Information Theory}, vol.~62, no.~4, pp. 1897--1910, 2016.

\bibitem{management}
A.~Kosta, N.~Pappas, A.~Ephremides, and V.~Angelakis, ``Age of information
  performance of multiaccess strategies with packet management,'' \emph{Journal
  of Communications and Networks}, vol.~21, no.~3, pp. 244--255, 2019.

\bibitem{deadlines}
C.~Kam, S.~Kompella, G.~D. Nguyen, J.~E. Wieselthier, and A.~Ephremides, ``On
  the age of information with packet deadlines,'' \emph{IEEE Transactions on
  Information Theory}, vol.~64, no.~9, pp. 6419--6428, 2018.

\bibitem{schedule_1}
I.~Kadota, A.~Sinha, E.~Uysal-Biyikoglu, R.~Singh, and E.~Modiano, ``Scheduling
  policies for minimizing age of information in broadcast wireless networks,''
  \emph{IEEE/ACM Transactions on Networking}, vol.~26, no.~6, pp. 2637--2650,
  2018.

\bibitem{schedule_2}
R.~Talak, S.~Karaman, and E.~Modiano, ``Improving age of information in
  wireless networks with perfect channel state information,'' \emph{IEEE/ACM
  Transactions on Networking}, vol.~28, no.~4, pp. 1765--1778, 2020.

\bibitem{schedule_4}
C.~Li, Q.~Liu, S.~Li, Y.~Chen, Y.~T. Hou, W.~Lou, and S.~Kompella, ``Scheduling
  with age of information guarantee,'' \emph{IEEE/ACM Transactions on
  Networking}, pp. 1--14, 2022.

\bibitem{schedule_5}
H.~Farag, M.~Gidlund, and {\v{C}}.~Stefanović, ``A deep reinforcement learning
  approach for improving age of information in mission-critical iot,'' in
  \emph{2021 IEEE Global Conference on Artificial Intelligence and Internet of
  Things (GCAIoT)}, 2021, pp. 14--18.

\bibitem{D2D_resource}
M.~Elnourani, S.~Deshmukh, and B.~Beferull-Lozano, ``Distributed resource
  allocation in underlay multicast d2d communications,'' \emph{IEEE
  Transactions on Communications}, vol.~69, no.~5, pp. 3409--3422, 2021.

\bibitem{D2D_traffic}
G.~Chisci, H.~Elsawy, A.~Conti, M.-S. Alouini, and M.~Z. Win, ``Uncoordinated
  massive wireless networks: Spatiotemporal models and multiaccess
  strategies,'' \emph{IEEE/ACM Transactions on Networking}, vol.~27, no.~3, pp.
  918--931, 2019.

\bibitem{D2D_fairness}
M.~Liu and L.~Zhang, ``Resource allocation for d2d underlay communications with
  proportional fairness using iterative-based approach,'' \emph{IEEE Access},
  vol.~8, pp. 143\,787--143\,801, 2020.

\bibitem{D2D_AoI_0}
P.~D. Mankar, M.~A. Abd-Elmagid, and H.~S. Dhillon, ``Spatial distribution of
  the mean peak age of information in wireless networks,'' \emph{IEEE
  Transactions on Wireless Communications}, vol.~20, no.~7, pp. 4465--4479,
  2021.

\bibitem{D2D_AoI_1}
H.~H. Yang, A.~Arafa, T.~Q.~S. Quek, and H.~V. Poor, ``Optimizing information
  freshness in wireless networks: A stochastic geometry approach,'' \emph{IEEE
  Transactions on Mobile Computing}, vol.~20, no.~6, pp. 2269--2280, 2021.

\bibitem{D2D_AoI_2}
J.~Li, D.~Wu, C.~Yue, Y.~Yang, M.~Wang, and F.~Yuan, ``Energy-efficient
  transmit probability-power control for covert d2d communications with age of
  information constraints,'' \emph{IEEE Transactions on Vehicular Technology},
  pp. 1--15, 2022.

\bibitem{D2D_AoI_3}
M.~Li, C.~Chen, H.~Wu, X.~Guan, and X.~Shen, ``Age-of-information aware
  scheduling for edge-assisted industrial wireless networks,'' \emph{IEEE
  Transactions on Industrial Informatics}, vol.~17, no.~8, pp. 5562--5571,
  2021.

\bibitem{schedule_3}
I.~Kadota, A.~Sinha, and E.~Modiano, ``Scheduling algorithms for optimizing age
  of information in wireless networks with throughput constraints,''
  \emph{IEEE/ACM Transactions on Networking}, vol.~27, no.~4, pp. 1359--1372,
  2019.

\bibitem{D2D_AoI_5}
J.~Sun, L.~Wang, Z.~Jiang, S.~Zhou, and Z.~Niu, ``Age-optimal scheduling for
  heterogeneous traffic with timely throughput constraints,'' \emph{IEEE
  Journal on Selected Areas in Communications}, vol.~39, no.~5, pp. 1485--1498,
  2021.

\bibitem{D2D_AoI_4}
L.~Luo, Z.~Liu, Z.~Chen, M.~Hua, W.~Li, and B.~Xia, ``Age of information-based
  scheduling for wireless d2d systems with a deep learning approach,''
  \emph{IEEE Transactions on Green Communications and Networking}, pp. 1--1,
  2022.

\bibitem{AoI_Delay_multiserver1}
A.~M. Bedewy, Y.~Sun, and N.~B. Shroff, ``Minimizing the age of information
  through queues,'' \emph{IEEE Transactions on Information Theory}, vol.~65,
  no.~8, pp. 5215--5232, 2019.

\bibitem{AoI_Delay_multiserver2}
R.~Talak and E.~H. Modiano, ``Age-delay tradeoffs in queueing systems,''
  \emph{IEEE Transactions on Information Theory}, vol.~67, no.~3, pp.
  1743--1758, 2021.

\bibitem{AoI_Delay_1}
R.~Devassy, G.~Durisi, G.~C. Ferrante, O.~Simeone, and E.~Uysal, ``Reliable
  transmission of short packets through queues and noisy channels under latency
  and peak-age violation guarantees,'' \emph{IEEE Journal on Selected Areas in
  Communications}, vol.~37, no.~4, pp. 721--734, 2019.

\bibitem{poission}
P.~Park, S.~Coleri~Ergen, C.~Fischione, C.~Lu, and K.~H. Johansson, ``Wireless
  network design for control systems: A survey,'' \emph{IEEE Communications
  Surveys Tutorials}, vol.~20, no.~2, pp. 978--1013, 2018.

\bibitem{Ack_free}
R.~Talak, S.~Karaman, and E.~Modiano, ``Optimizing age of information in
  wireless networks with perfect channel state information,'' in \emph{2018
  16th International Symposium on Modeling and Optimization in Mobile, Ad Hoc,
  and Wireless Networks (WiOpt)}, 2018, pp. 1--8.

\bibitem{PPP}
M.~Haenggi, \emph{Geometry for Wireless Networks}.\hskip 1em plus 0.5em minus
  0.4em\relax Cambridge, UK: Cambridge Univ. Press, 2012.

\bibitem{arrival}
E.~T. Ceran, D.~Gündüz, and A.~György, ``Average age of information with
  hybrid arq under a resource constraint,'' \emph{IEEE Transactions on Wireless
  Communications}, vol.~18, no.~3, pp. 1900--1913, 2019.

\bibitem{m1}
I.~Kadota, A.~Sinha, E.~Uysal-Biyikoglu, R.~Singh, and E.~Modiano, ``Scheduling
  policies for minimizing age of information in broadcast wireless networks,''
  \emph{IEEE/ACM Transactions on Networking}, vol.~26, no.~6, pp. 2637--2650,
  2018.

\bibitem{m2}
B.~Sombabu and S.~Moharir, ``Age-of-information based scheduling for
  multi-channel systems,'' \emph{IEEE Transactions on Wireless Communications},
  vol.~19, no.~7, pp. 4439--4448, 2020.

\bibitem{capture}
M.~Zorzi and R.~Rao, ``Capture and retransmission control in mobile radio,''
  \emph{IEEE Journal on Selected Areas in Communications}, vol.~12, no.~8, pp.
  1289--1298, 1994.

\bibitem{24}
L.~Kleinrock, \emph{{Queueing Systems, Volume 1: Theory}}.\hskip 1em plus 0.5em
  minus 0.4em\relax London, UK: Wiley-Interscience, 1975.

\bibitem{tx_time}
D.~P. Bertsekas and R.~G. Gallager, \emph{{Data Networks}}.\hskip 1em plus
  0.5em minus 0.4em\relax NJ, USA: Prentice-Hall, 1992.

\bibitem{thinning}
M.~Haenggi, \emph{{Stochastic Geometry for Wireless Networks}}.\hskip 1em plus
  0.5em minus 0.4em\relax Cambridge, UK: Cambridge Univ. Press, 2012.

\bibitem{first_order_opt}
I.~Griva, S.~Nash, and A.~Sofer, \emph{{Linear and Nonlinear
  Optimization}}.\hskip 1em plus 0.5em minus 0.4em\relax Cambridge, UK:
  Cambridge Univ. Press, 2009.

\bibitem{Laplace}
M.~Haenggi and R.~K. Ganti, \emph{Interference in Large Wireless
  Networks}.\hskip 1em plus 0.5em minus 0.4em\relax Boston, USA: Now
  Foundations and Trends, 2009.

\bibitem{approx_expec}
N.~Lee, X.~Lin, J.~G. Andrews, and R.~W. Heath, ``Power control for d2d
  underlaid cellular networks: Modeling, algorithms, and analysis,'' \emph{IEEE
  Journal on Selected Areas in Communications}, vol.~33, no.~1, pp. 1--13,
  2015.

\bibitem{Loynes}
R.~M. Loynes, ``The stability of a queue with non-independent interarrival and
  service times,'' \emph{Math. Proc. Cambridge Philos. Soc}, vol.~58, no.~3, p.
  497–520, 1962.

\end{thebibliography}
	
\end{document}